\documentclass[12pt]{article}
\usepackage{epsfig,equations,graphics,amssymb}
\usepackage{cite}

\begin{document}

\def\nicefrac#1#2{\hbox{${#1\over #2}$}}
\textwidth 10cm
\setlength{\textwidth}{13.7cm}
\setlength{\textheight}{23cm}

\oddsidemargin 1cm
\evensidemargin -0.2cm
\addtolength{\topmargin}{-2.5 cm}

\newcommand{\nn}{\nonumber}
\newcommand{\raw}{\rightarrow}
\newcommand{\be}{\begin{equation}}
\newcommand{\ee}{\end{equation}}
\newcommand{\bea}{\begin{eqnarray}}
\newcommand{\eea}{\end{eqnarray}}
\newcommand{\dl}{\stackrel{\leftarrow}{D}}
\newcommand{\dr}{\stackrel{\rightarrow}{D}}
\newcommand{\dd}{\displaystyle}
\newcommand{\Ln}{{\rm Ln}}

\pagestyle{empty}

\begin{flushright}
FTUAM 01/13 \\
IFT-UAM/CSIC 01/19  \\
hep--ph/0106267 \\
June 2001 \\
\end{flushright}
\vskip1cm

\renewcommand{\thefootnote}{\fnsymbol{footnote}}
\begin{center}
{\Large\bf
Optimal observables to search for indirect Supersymmetric QCD signals in Higgs boson
 decays}\\[1cm]
{\large Ana M. Curiel, 
Mar\'{\i}a J. Herrero, David Temes \\ 
and Jorge F. de Troc\'oniz~\footnote{e-mail addresses: 
curiel@delta.ft.uam.es, herrero@delta.ft.uam.es, 
temes@delta.ft.uam.es, troconiz@mail.cern.ch}
}\\[6pt]
{\it Departamento de F\'{\i}sica Te\'{o}rica \\
   Universidad Aut\'{o}noma de Madrid,
   Cantoblanco, 28049 Madrid, Spain.}
\\[1cm]

\begin{abstract}
\vspace*{0.4cm}

 In this work we study the indirect effects of squarks and gluinos via 
 supersymmetric QCD radiative corrections in the decays of 
 Higgs particles within the Minimal Supersymmetric Standard Model. We consider
 a heavy supersymmetric spectrum and focus on the main nondecoupling effects.
 We propose a set of observables that are sensitive to these corrections and 
 that will be accessible at the CERN Large Hadron Collider and Fermilab Tevatron. These observables are the ratios of
 Higgs boson branching ratios into quarks  divided by the corresponding 
 Higgs boson branching ratios 
 into leptons, and both theoretical and experimental uncertainties are 
expected to be minimized.  We show that these nondecoupling
 corrections are sizable
 for all the proposed observables in the large $\tan\beta$ region and are
 highly correlated. A global analysis of all these observables 
 will allow the experiments to reach the highest sensitivity to indirect supersymmetric signals.
 
\end{abstract}

\end{center}

\vfill
\clearpage

\renewcommand{\thefootnote}{\arabic{footnote}}
\setcounter{footnote}{0}

\pagestyle{plain}

\section{Introduction} 

One of the main goals of the next generation of colliders will be to explore the 
Higgs sector phenomenology. 
The simplest candidates for Higgs sector
physics beyond the standard model (SM) are the two Higgs doublet models 
(2HDM)~\cite{2HDM}, and, among these, the leading one is provided by the minimal
supersymmetric standard model (MSSM)~\cite{MSSM}. This requires a Higgs sector 
called of type II, where one of the Higgs doublets couples  to the toplike
quarks and the other to the bottomlike quarks. The basic differences 
between the MSSM and a general 2HDM of type II (2HDMII) depend, first, on the 
supersymmetric (SUSY) sector, which is obviously absent in the 2HDMII, and, second, on the
different values of the Higgs boson self-couplings. In the MSSM, because of the
underlying supersymmetry, these couplings are fixed in terms of the electroweak 
gauge couplings. 

In all these models, the physical Higgs boson spectrum consists of two
neutral CP-even Higgs bosons $h^o$ and $H^o$, one CP-odd Higgs boson $A^o$,
and two charged Higgs bosons $H^{\pm}$~\cite{2HDM}.  The 
tree-level parameters
are the Higgs boson masses $m_{h^o}$, $m_{H^o}$, $m_{A^o}$, and $m_{H^{\pm}}$,
the mixing angle in the CP-even neutral sector, $\alpha$, and the ratio of the vacuum 
expectation values of the two Higgs doublets, $\tan \beta =v_2/v_1$.
In the MSSM, due to supersymmetry, these can be written in terms of
just two 
free parameters, which are
usually chosen as $m_{A^o}$ and  $\tan \beta$. The Higgs boson masses are therefore 
not independent parameters in the MSSM, as they are in a general 2HDMII. Once
the radiative corrections are included these tree-level mass relations are
significantly modified, but there still remains a clear pattern for the Higgs boson
masses in terms of the MSSM parameters~\cite{loopmassMSSM,pierce}. 
Hence, if all these masses could be
measured with good precision the mass pattern itself would be a first 
indirect indication of their SUSY origin. 

The most striking prediction of the MSSM is that the lightest 
Higgs boson mass lies below 130-135 GeV, within the reach of   
present and next generation colliders, 
the Fermilab Tevatron~\cite{TevRunII} and the CERN Large Hadron Collider (LHC)~\cite{lhc}. 
 The properties of this Higgs boson $h^o$, however, are 
very similar to those of the SM Higgs particle in the so-called decoupling regime 
where $m_{A^o}\gg m_Z$~\cite{decoupling}.
For the relevant $h^o$ production processes and decay channels at the LHC, this
decoupling already  occurs, in practice, at $m_{A^o}$ values not far from the 
electroweak scale $m_{\rm EW}$. For instance, the difference in the width 
of the main decay
$h^o\rightarrow b \bar b$, from the SM value is less than $10\%$ for 
$m_ {A^o}> 350 \,{\rm GeV}$ and $2<\tan \beta<50$. 
There is just one region, corresponding to low $m_{A^o}$ and large 
$\tan \beta$ values, where the  pattern of 
$h^o$ branching ratios significantly differs from the SM one. It is because 
in this region the dominant 
decays to $b\bar b$ pairs and to $\tau^+ \tau^-$ pairs are enhanced and, as a
consequence, the subdominant channels are significantly suppressed. 
Thus, for
most of the ($m_ {A^o}$, $\tan \beta$) parameter space it seems difficult
to disentangle the nonstandard nature of $h^o$ 
and, in order to decide on 
its
SUSY origin, one must consider in addition the phenomenology of the
other Higgs particles (for a review see, for instance, \cite{spiraQCD}). 
The relevant question then becomes one of finding SUSY signals by looking into 
$h^o$, $H^o$, $A^o$ and $H^{\pm}$ production and decays at colliders, and of
distinguishing these signals from those of a general 2HDMII. Obviously, if the SUSY 
particles are within the reach of the LHC and/or the Tevatron, the priority is to look for their direct
production. However, it may well happen that these turn out 
to be too heavy to be produced directly. In that case one will have to 
look for
indirect SUSY signals via their contributions as virtual intermediate states in 
standard processes, in much the same way as was done in the
past at the CERN $e^+e^-$ collider LEP for indirect effects from top and Higgs
particles
in  precision observables. 
We are particularly
interested here in their contributions to the radiative corrections involved in
Higgs physics. 

In this work, we consider a  scenario where some or all of
the Higgs particles have been discovered and their masses and branching ratios 
have been measured, but the SUSY particles have not shown up yet. We will study 
optimal strategies to look for indirect SUSY signals via their
radiative corrections in Higgs boson decays~\cite{Dabelstein,cjs,solaHTB,EberlHTB,solaEW,topQCD,topEW}.  
Since it is known that
there are specific decay processes where some SUSY radiative corrections do not
decouple, even in the case of a extremely heavy SUSY spectrum
~\cite{cmwpheno,polonsky,kolda,HaberTemes,CarenaDavid,EberlTodos,ourHtb,dobado}, we will
concentrate on these particular channels. This nondecoupling behavior is a genuine part of Higgs 
sector physics and has not been found yet in other MSSM sectors, such as, for instance,
electroweak gauge boson physics~\cite{TesisS}. The nondecoupling processes we consider here
are the Higgs boson decay channels 
into quark pairs, and more specifically we will concentrate on those that are
enhanced at large $\tan \beta$ values: 
$h^o\rightarrow b \bar b$,    
$H^o\rightarrow b \bar b$, $A^o\rightarrow b \bar b$~\cite{Dabelstein,polonsky,kolda,cjs,EberlTodos,
cmwpheno,HaberTemes} and 
$H^+\rightarrow t \bar b$~\cite{solaHTB,EberlHTB,CarenaDavid,solaEW,EberlTodos,ourHtb}.
We will also analyze the top quark decay channel 
$t \rightarrow H^+ b$~\cite{topQCD,topEW} which is complementary to the 
$H^+\rightarrow t \bar b $ decay for low 
$m_{A^o}$ values. We will focus, in particular, on the SUSY QCD corrections 
from third-generation squarks, $\tilde q$, and gluinos, $\tilde g$, which are 
the dominant SUSY radiative corrections for most of 
the MSSM parameter space and are numerically sizable~\cite{Dabelstein,cjs,solaHTB,EberlHTB,topQCD}. 
For
instance, in $H^+\rightarrow t \bar b$ they can be as large as $60\%$ and of
either sign for 
quasidegenerate gluinos and squarks of mass 1 TeV and $\tan\beta =50$. These 
nondecoupling corrections were derived recently in~\cite{dobado}
for all the Higgs boson decay channels into all
possible quark pairs, and analyzed in more detail
for the 
$h^o \rightarrow b \bar b$~\cite{cmwpheno,polonsky,kolda,EberlTodos,HaberTemes}
 and $H^+\rightarrow t \bar b$ ~\cite{CarenaDavid,EberlTodos,ourHtb}
channels. 
For nearly degenerate gluinos and squarks these SUSY QCD
corrections, to one-loop level, can be generically written as 
$\Gamma = \Gamma^0(1+\alpha_S\frac{\mu M_{\tilde g}}{M_{\rm SUSY}^2}K)$, 
where $\Gamma^0$ is the corresponding partial decay width without the SUSY QCD
contribution, $\alpha_S$ is the strong coupling constant, $\mu$ is the 
bilinear Higgs boson MSSM parameter, $M_{\tilde g}$ is the gluino mass, 
$M_{\rm SUSY}$ is
a common generic SUSY mass characterizing the squark masses, and $K$ 
is a function of $\beta$ and $\alpha$ that depends on the specific 
channel~\cite{dobado}. The nondecoupling behavior is seen here as a nonvanishing
contribution to the partial width of order ${\cal O}(\alpha_S K)$, which is
present even in the limit of a heavy SUSY spectrum where 
$|\mu| \sim M_{\tilde g} \sim M_{\rm SUSY}\gg m_{\rm EW}$.  
These corrections are known to be particularly relevant for the    
$h^o, H^o, A^o \rightarrow b \bar b$, $H^+\rightarrow t \bar b$, and 
$t \rightarrow H^+ b$ decays,
because it is precisely in these channels that
the function $K$ grows linearly with $\tan \beta$ and can provide
sizable contributions for the interesting large $\tan \beta$ region. 
It is worth  mentioning  that the SUSY electroweak (SUSY EW) corrections may also be 
relevant in particular regions of the MSSM parameter 
space~\cite{Dabelstein,solaEW,topEW}. In the 
large $\tan \beta$ regime and for quasidegenerate squarks, neutralinos, and
charginos, these are known to behave as 
$\sim (\frac{h_t}{4\pi})^2\frac{\mu A_t}{M_{\rm SUSY}^2}\tan \beta$, 
where $M_{\rm SUSY}$ is the common generic mass
for squarks, neutralinos, and charginos, $h_t$ is the tree-level top quark Yukawa 
coupling, $h_t=\frac{gm_t}
{{\sqrt 2}m_W \sin \beta}$, and $A_t$ is the top quark trilinear 
coupling~\cite{cmwpheno,polonsky,kolda}. For small or moderate $A_t$ values and a heavy
SUSY spectrum, these SUSY EW corrections are considerably smaller than
the SUSY QCD ones and
can be ignored, but for very large $A_t$ values, $|A_t| \sim |\mu| \sim M_{\rm SUSY} \gg m_{\rm EW}$, these SUSY EW corrections do not decouple either and, depending on their sign, can increase or decrease the nondecoupling effect of the SUSY QCD corrections.
             
In order to select optimal observables to look for these indirect 
SUSY QCD signals we require of them, first, to be measurable at the 
Tevatron~\cite{TevRunII},  
LHC~\cite{lhc}, and/or the next generation linear colliders;\footnote{The case of linear 
colliders is not explicitly discussed here, but most of our results apply to them as 
well. This case has been considered recently in~\cite{HaberLC}.}.
 second, to be most 
sensitive to the mentioned SUSY QCD 
corrections;  and, third, to have the minimal theoretical and experimental 
uncertainties. We propose
and analyze here a set of observables that satisfy all these conditions. 
These are the ratios of the branching ratios of the Higgs boson decays into third-generation 
quarks to the branching ratios of this same Higgs boson into 
third-generation leptons\footnote{Some preliminary results 
were presented by one of us in~\cite{MJsitges}.}. 
We will show that these ratios can be considered optimal observables because of the following reasons: \\ 
$\bullet$ The SUSY QCD nondecoupling corrections contribute just to the numerator,
 that is, to quark decays and not to lepton decays. The first will be considered 
 as the search channel, the second as
the control channel.
 \\
$\bullet$  These corrections are
 maximized at large $\tan \beta$ and are sizable enough to be 
 measurable. \\
$\bullet$  The production uncertainties are minimized in ratios.  \\
$\bullet$  They will be experimentally accessible at the LHC or Tevatron. \\
$\bullet$  They will allow one
 to distinguish the MSSM Higgs sector from a general 2HDMII. 

 The paper is organized as follows.  Section 2 is devoted to an
 experimental overview of Higgs particle searches at the Tevatron and LHC. 
 The relevant regions in the $(m_{A^o},\tan\beta)$ plane are briefly reviewed.
 In Section 3 the set of optimal observables is presented and analyzed in full 
 detail. The leading contributions from tree-level and standard QCD
 corrections to these observables are also studied and their theoretical 
 uncertainties estimated. The explicit formulas for nondecoupling SUSY QCD corrections 
 from heavy squarks and gluinos are presented. A discussion of the relevance of the SUSY EW corrections is also contained in Section 3. A comparison between a general
 2HDMII and the MSSM is also included. The numerical results for the 
 proposed observables and a discussion as a function of the MSSM parameters are 
 presented in Section 4. Finally, the conclusions are summarized in Section 5.


\section{Experimental overview}
 
In this section we turn to the perspectives of the experimental
measurement itself. We focus on the next hadron 
colliders: the Tevatron and LHC.
From this point of view, a measurement of the SUSY radiative corrections
in Higgs physics is the next
step in the problem of discovering a signal of the existence of one or
several Higgs particles. Only after a discovery, but immediately after, will
the question of what is the underlying structure of the Higgs sector be
raised. As we have said, this is the principal motivation of our study.

The discovery reach problem has been studied carefully and extensively by 
several working groups, both for the Tevatron Run 2~\cite{TevRunII} and for the 
LHC~\cite{lhc}.
In particular, the searches for SM or MSSM Higgs bosons were used as  
benchmark channels during the design of the ATLAS and CMS 
detectors.

Following the results of these studies, one learns that for values of
$\mbox{tan}~\beta$ large enough so that the corrections are sizable, there
are three regions in the ($m_{A^o}$, $\mbox{tan}~\beta$) plane of phenomenological 
interest.

The first is the region of large values of $\mbox{tan}~\beta$ 
(say $\gtrsim 10$), and low $m_{A^o}$ (say
$m_A \lesssim 120$ GeV). Here several production processes grow with
$\mbox{tan}^2~\beta$, and the corresponding cross sections are much larger
than the typical SM ones. One example is the associated production of a
CP-odd boson $A^o$ and bottom quarks: $gg (q\bar{q}) \to b\bar{b} A^o$.
In this region of the MSSM, the lighter CP-even boson $h^o$ has essentially
the same couplings and mass (within 5\%) as the $A^o$, effectively doubling
the production cross section~\cite{2HDM}. The heavier CP-even boson 
$H^o$ has SM bosonlike couplings, and its production rates are considerably smaller. In turn,
the charged Higgs boson $H^{\pm}$ is lighter than the top quark, and $t \to b H^+$ 
happens in a sizable part of the top quark decays.

With such large cross sections and relatively small masses this has been 
the main region examined at the Tevatron during Run 1.
In fact, the Collider Detector at Fermilab (CDF) Collaboration has already searched for associated production 
of $b\bar{b} A^o(h^o)$, where $A^o(h^o) \to b\bar{b}$~\cite{juan}, or 
$A^o(h^o) \to \tau^+ \tau^-$~\cite{tom}, excluding a part of the 
($m_{A^o}$, $\tan \beta$) plane. 
Also, the CDF and D0 Collaborations have excluded a part of the MSSM parameter space 
corresponding to unobserved decays $t \to b H^+$, with $H^+ \to \tau^+ \nu$
~\cite{chiggs}, more specifically, the region where
$\frac{\Gamma(t\to H^+ b)}{\Gamma(t\to W^+ b)} \gtrsim 1$.
It is clear that the search will continue in Run 2,
and a discovery is possible before the start of the LHC~\cite{TevRunII}.

The first analysis takes advantage  of both the large cross section 
(${\cal O}$(1-100 pb)) and large $\mbox{B}(A^o \to b\bar{b}) \sim 90\%$. 
The experimental signature is
four jets with at least three of them $b$ tagged. This channel has also
the advantage that it allows the reconstruction of $m_{A^o}$ and, once this mass
is known, the production rate gives an idea of the size of $\mbox{tan}~\beta$.
On the other hand, as in any search in a purely hadronic channel, the QCD 
background has to be properly understood. Systematic errors at the $20-30\%$
level are not infrequent~\cite{juan}.

In contrast, the channel $A^o \to \tau^+ \tau^-$, has to overcome a relatively 
smaller branching ratio ($\sim$10\%). 
However, the electroweak background $Z (\to \tau^+ \tau^- )+{\rm jets}$ is
considerably less dangerous~\cite{tom}.
 
If an $A^o$ signal were found in these analyses, the corresponding ratio of 
$b\bar b$ to $\tau^+ \tau^-$
events would be a direct test of the SUSY QCD corrections.

A possible signal in the charged Higgs boson channel could also be 
useful~\cite{topEW}. In this case, it also happens that the 
SUSY QCD corrections do not decouple in  the $t \to b H^+$ decay but they do
decouple in the standard channel, $t \to b W^+$~\cite{MJsitges,RADCOR2000}. 
This is the reason why  $\Gamma(t \to b W^+)$ is considered now as 
the corresponding control width.
The problem in this case is that the tree-level prediction
changes rapidly with $m_{H^+}$ and $\mbox{tan}~\beta$. 
On the other hand, the accuracy in 
the experimental reconstruction of $m_{H^+}$ in the decay $H^+ \to \tau^+ \nu$
is limited, and this channel by itself does not provide information about
$\mbox{tan}~\beta$. It should be pointed out that the LHC experiments will 
be able to explore this channel very efficiently.

The second region of interest still has large $\mbox{tan}~\beta$, but
$140$ GeV $\lesssim m_{A^o} 
\lesssim 500$ GeV\footnote{The properties of the Higgs sector in 
the intermediate region $120$ GeV $\lesssim m_{A^o} \lesssim 140$ GeV
are a mixture of the 
two cases discussed. A detailed description is complicated and beyond the 
scope of this paper.}.
 With such large masses this can be considered the genuine search zone of 
the LHC experiments~\cite{lhc}.
In this zone, the heavier CP-even boson $H^o$ has 
similar
couplings and mass than the $A^o$ boson, again doubling the cross section of the process
$gg (q\bar{q}) \to b\bar{b} A^o(H^o)$ (${\cal O}$(10-1000 pb)). Now the light CP-even Higgs boson $h^o$ 
behaves like the one in the SM with $m_{h^o} \sim 110-130$ GeV (the precise mass
value depends on the choice of the MSSM parameters).
In this region, the mass of the charged Higgs boson is either very close to
$m_t$ and $\mbox{B}(t \to bH^+)$ is negligible, or $m_{H^+} > m_t$ and
$t \to b H^+$ is directly forbidden. Then, the main possibility of producing
charged Higgs bosons is in the process $gg (q\bar{q}) \to b\bar{t} H^+$. Again, this
cross section grows approximately (at large enough $\tan\beta$) with $\mbox{tan}^2~\beta$. 
                
The main possibility of the Tevatron here 
would be to detect the SM-like boson $h$ in the associated production 
$q\bar{q} \to W(Z)\ h^o$ and $h^o \to b\bar{b}$. Also, the analysis of the 
ratio $(A^o \to b\bar{b})/(A^o \to \tau^+ \tau^-)$ can be extended, but only 
for relatively small values of $m_{A^o}$ and very large values of 
$\mbox{tan}~\beta$~\cite{TevRunII}.

With respect to these last channels at the LHC, the challenge will be to detect
the decay $A^o(H^o) \to b\bar{b}$ in events with four $b$ jets. Studies have shown 
that to control the QCD background will be very difficult. In contrast, it will be 
possible to measure the decay $A^o(H^o) \to \tau^+ \tau^-$ and even 
$A^o(H^o) \to \mu^+ \mu^-$. Both channels are leptonic, and therefore do not 
allow an estimation of the SUSY QCD corrections. However, they will provide
good measurements of $m_{A^o}$ and $\mbox{tan}~\beta$, which can be used in 
addition to other analyses (say, $H^+ \to \tau^+ \nu$). Current studies for the 
LHC~\cite{lhc} indicate that values
of $\Delta m_{A^o}/m_{A^o} $ better than $\sim 1\%$ and of 
$\Delta\tan\beta/\tan\beta$
better than $\sim 12\%$ ( if $\tan\beta>10$) can be reached.  
 
In the case that $m_{H^+} > m_t + m_b$, the decay channel 
$H^+ \to t\bar{b}$ opens and can be used not only to complement the decay 
$H^+ \to \tau^+ \nu$, but also to provide an independent estimation of
$m_{A^o}$ (in the frame of the MSSM) and $\mbox{tan}~\beta$.

Finally, we mention the very interesting region with $m_{A^o} \gtrsim 140$ GeV and
intermediate $3 \lesssim \mbox{tan}~\beta \lesssim 10$.  At the LHC this zone is called
the ``hole'' because the only Higgs particle accessible is the SM-like one.
Here the main question would be not only to discriminate the MSSM from a 
more general 2HDM sector, but to know if there is something beyond the plain
SM. 

We will see in the next sections how the SUSY QCD corrections can be used
for that matter.


\section{Optimal observables}

In this section we present the set of observables that are the most sensitive 
to the nondecoupling SUSY QCD corrections.

Since we are looking for indirect signals of SUSY in 
an experimental scenario described by the discovery of Higgs particles, with 
masses following the MSSM pattern, and where the SUSY particles are 
too heavy to be produced in colliders, we will search for observables with 
very specific conditions. From the theoretical point of view the requirements 
are the following: 
\begin{itemize}
\item In order to discriminate between the MSSM and a nonsupersymmetric 2HDMII, 
we need observables with different predictions for these two models.
\item These predictions should be distinguishable even in the case of a very 
heavy SUSY spectrum.
\item The theoretical uncertainties must be minimized. Some specific 
ratios of branching ratios will cancel  these
uncertainties either totally or partially. In particular, we wish the $\tan\beta$  and $\alpha$ 
dependence to appear just 
in the correction, but not in the lowest order contribution.
\end{itemize}

On the other hand, the experimental requirements are the following:
\begin{itemize}
\item A control channel is needed in order to eliminate systematic errors, 
so that  
ratios of event rates are better than event rates themselves.
\item The corrections to the observables should be 
sizable in the large $\tan \beta$ regime, since the relevant Higgs particle
production cross sections grow with this 
parameter, such that we have better statistics for larger values 
of $\tan \beta$.
\item There should be good identification of the final state particles. 
\item Experimental uncertainties from Higgs particle production must be
minimized. 
These will cancel in some specific ratios of events from Higgs particle decays.
\item The expected accuracy at the LHC and Tevatron for these observables should be
 good enough that the deviation produced by the corrections becomes
 apparent. In particular, the size of the SUSY QCD corrections
 must be larger than the expected error bars.  
\end{itemize}

We propose the following set of optimal observables, which satisfy the previous
theoretical and experimental requirements: 

\begin{eqnarray}
O_{h^o} \equiv \frac{B(h^o \rightarrow b\bar b)}{B(h^o \rightarrow \tau^+ \tau^-)}
\, \, &,& \, \, 
O_{H^o} \equiv \frac{B(H^o \rightarrow b\bar b)}{B(H^o \rightarrow \tau^+ \tau^-)}, 
\nn \\
O_{A^o} \equiv \frac{B(A^o \rightarrow b\bar b)}{B(A^o \rightarrow \tau^+ \tau^-)}
 \, \, &,& \, \, 
O_{H^+} \equiv \frac{B(H^+ \rightarrow t\bar b)}{B(H^+ \rightarrow \tau^+ \nu)}. 
\nn \\
\label{eq.interobsH}
\end{eqnarray}

In addition, we consider the following ratio:  
\begin{equation}
O_{t} \equiv \frac{B(t \to H^+ b)}{B(t \to W^+ b)}, 
\label{eq.interobst}
\end{equation}
which
complements the charged Higgs boson observable in the low $m_{A^o}$ region.
The predictions for these observables at the tree level are the same for  
any general 2HDMII,
and in particular for the MSSM. These are

\begin{eqnarray}
&O_{\phi}^{\rm tree} = N_c \frac{m_b^2 \tilde \beta_b^3}{m_{\tau}^2 \tilde \beta_{\tau}^3}
\, \, \, , \, \, \,
\phi = h^o, H^o \, ; \qquad
O_{A^o}^{\rm tree} = N_c \frac{m_b^2 \tilde \beta_b}
{m_{\tau}^2 \tilde \beta_{\tau}} \, ;& \nn \\
&O_{H^+}^{\rm tree}= N_c m^2_{H^+}\frac{(m^2_{H^+}-m_t^2-m_b^2)
(m_t^2 \cot^2\beta +
 m_b^2 \tan^2\beta)-4m_t^2m_b^2}
{m_{\tau}^2 \tan^2\beta (m^2_{H^+}-m_{\tau}^2)^2}
\lambda^{1/2}_{H^+,t,b} \, ;
\nn \\
&O_t^{\rm tree} = 
\frac{(m_t^2 + m_b^2 - m^2_{H^+}) (m_t^2 \cot^2\beta + m_b^2 \tan^2\beta) 
+ 4 m_b^2 m_t^2}{ (m_t^2 + m_b^2 - 2M^2_{W^+})M^2_{W^+} + 
(m_t^2 - m_b^2)^2} \, \, 
\nicefrac{\lambda^{1/2}_{t,H^+,b}}
{\lambda^{1/2}_{t,W^+,b}} \, ; &\nn \\
%
%
%
\label{eq.treeobserv}
\end{eqnarray}

where 
\begin{eqnarray}
\tilde \beta_f = \sqrt{1-4 m_f^2/M_{\phi}^2} \, \, &,& \, \,
\lambda^{1/2}_{x,y,z} = 
\sqrt{(1-(m_y^2+m_z^2)^2/m_x^4)(1-(m_y^2-m_z^2)^2/m_x^4)}. \nn
\end{eqnarray}

As a general remark,   
the differences in the predictions from the various models for all these
observables come in
the corrections beyond the tree level and will
be discussed in the following subsections.
In particular, it is interesting to emphasize that the predictions for the 
neutral
Higgs boson observables at the tree level
are independent of $\tan\beta$ and $\alpha$. The charged Higgs boson observable depends on
$\tan\beta$, but this dependence is very mild in the large  
$\tan\beta$ region. 
Therefore, these observables  
are especially sensitive to any extra contribution growing with 
$\tan\beta$. In the top quark decay observable, however, the tree level prediction 
goes as $\tan^2\beta$, for large $\tan\beta$, and it will be more difficult
to disentangle the mentioned contributions. Other analyses of the observables 
for the neutral channels in the large $\tan\beta$ limit and in the zero external
momentum approximation can be found in~\cite{kolda,siannah}. The neutral channel case with 
Higgs boson mass corrections included has been analyzed in~\cite{polonsky}.

\subsection{Supersymmetric QCD contributions} 
The nondecoupling SUSY QCD contributions to these observables in the MSSM 
can be easily
derived from the results for the effective Yukawa interactions of the  
Higgs sector with top and bottom quarks. These have been recently obtained 
at one-loop level, for arbitrary $\tan\beta$, and by a
functional integration of bottom and top squarks and gluinos 
in~\cite{dobado}. For nearly degenerate heavy squarks and gluinos, 
these corrections can be written as
 
\begin{eqnarray}
O_{h^o} &= O_{h^o}^o & 
\left[ 1- \frac{2\alpha_S}{3 \pi} \frac{M_{\tilde g} \mu}{M_{\rm SUSY}^2}
(\tan \beta +\cot \alpha)  \right] 
\nn \\
O_{H^o} &= O_{H^o}^o &
\left[ 1- \frac{2\alpha_S}{3 \pi} \frac{M_{\tilde g} \mu}{M_{\rm SUSY}^2}
(\tan \beta - \tan \alpha) \right] 
\nn \\
O_{A^o} &=  O_{A^o}^o &
\left[ 1- \frac{2\alpha_S}{3 \pi} \frac{M_{\tilde g} \mu}{M_{\rm SUSY}^2}
(\tan \beta + \cot \beta) \right] 
\nn \\
O_{H^+} &= O_{H^+}^o &
\left[ 1- \frac{2\alpha_S}{3 \pi} \frac{M_{\tilde g} \mu}{M_{\rm SUSY}^2}
(\tan \beta + \cot \beta) \right]  \nn \\
O_{t}&= O_{t}^o &
\left[ 1- \frac{2\alpha_S}{3 \pi} \frac{M_{\tilde g} \mu}{M_{\rm SUSY}^2}
(\tan \beta + \cot \beta) \right]. 
\label{eq.exprobs}
\end{eqnarray}

\vskip 0.2cm
Here, $\mu$ and $M_{\tilde g}$ are the MSSM bilinear parameter and the gluino mass,
respectively. $M_{\rm SUSY}$ is the common SUSY mass for squarks 
and gluinos ($M_{\rm SUSY}\sim M_{\tilde q}\sim M_{\tilde g}$) and  $\alpha _S$ is
the strong coupling constant evaluated at the corresponding decaying particle
mass.  
The leading terms $O^o$ refer to the value of the observables 
without the SUSY particle contributions and will be discussed in 
subsection~3.3. These formulas are for heavy squarks and gluinos, that is, 
for large $M_{\rm SUSY}$, and are valid for all $\tan \beta$ and $m_{A^o}$ values. 
Corrections to the previous formulas
are either of higher order in $\alpha_S$ or  
suppressed by higher inverse powers of the heavy SUSY masses 
$M_{\rm SUSY}$ and, for the present analysis, can be safely ignored. 
 
From a first look at eq.({\ref{eq.exprobs}}) one sees clearly that the 
observables have different predictions within the MSSM and in a general 2HDMII,
since the former includes the SUSY QCD corrections, which do not exist in 
the latter. This main
 difference comes from the nondecoupling behavior of the SUSY QCD corrections in the
 Higgs boson decays into quarks (and in the top quark decay into charged Higgs bosons),  
 which is relevant even for a very heavy SUSY spectrum, such that 
 $M_{\tilde g} \sim |\mu| \sim M_{\rm SUSY} \gg m_{\rm EW}$. Notice also that, 
 as mentioned
 before, 
these SUSY QCD contributions grow linearly with $\tan \beta$ so we expect
sizable corrections for large $\tan \beta$.

There are two limiting situations that are worth mentioning: 
the large $m_{A^o}$ limit $m_{A^o}\gg m_Z$,  and the large 
$\tan\beta$ limit $\tan\beta \gg 1$. In the former limit, $\cot\alpha$
approaches $-\tan\beta$ (correspondingly, 
$\tan\alpha \rightarrow -\cot\beta$) and, as can be seen 
in eq.({\ref{eq.exprobs}}), the SUSY QCD correction decouples in $O_{h^o}$. 
The other observables get exactly the same nondecoupling 
correction, proportional to $(\tan\beta+\cot\beta)$. In addition, 
in this limit,
the $H^o$ and $H^+$ masses approach the large $A^o$ mass. 
The leading contribution $O^o_{h^o}$ approaches the SM prediction and, 
therefore, 
as noted before, the situation cannot be distinguished from the
SM case. In the large $\tan \beta$ limit, the corrections do not decouple in 
any channel, they are all proportional to $\tan\beta$ and all have the same sign.
The universal character of the corrections will be very useful in a
global analysis, because it will yield strongly correlated signals
 in all the channels.  

\subsection{Other nondecoupling contributions in the minimal supersymmetric standard model}

It is well known that, in particular regions of the MSSM parameter
space the  
radiative corrections from the
SUSY EW sector can also be relevant in the decays of the
Higgs bosons and top quark that we
are considering here~\cite{Dabelstein,solaEW,topEW}. These
SUSY EW corrections are largely dominated by the
contributions with charginos or neutralinos
and third-generation sfermions in the loops; the contributions from
the Higgs sector
being negligible. For instance, in the $H^{+} \rightarrow t \bar b$
case with $m_{H^+} =250$ GeV and for light top and bottom squarks and light
charginos and neutralinos, $M_{\tilde q}, M_{\tilde \chi} \leq 200$ GeV,
they contribute as much as $\pm 20\%$ 
with respect to the tree level width, whereas the SUSY QCD
loops induced by squarks and gluinos are by far the leading SUSY
effects and give a contribution larger than $\pm 50\%$~\cite{solaEW}.
These SUSY EW corrections can only be competitive with the SUSY QCD
ones in the particular region of the SUSY parameter space where
$\tan\beta$ is very large ($\tan\beta > 20$), the bottom squarks are very
heavy ($M_{\tilde b} > 300$ GeV), and the top squarks and charginos are
relatively light ($M_{\tilde t}, M_{\tilde \chi} \sim 100-200$
GeV). For this particular choice, the total SUSY correction remains
around (30-50)$\%$ of the tree level width, with at most half of it of SUSY EW origin~\cite{solaEW}.

For our present assumption
of a very heavy SUSY spectrum with nearly degenerate
SUSY particle masses ($M_{\tilde t} \sim M_{\tilde b} \sim M_{\tilde
g} \sim M_{\tilde \chi} \sim |\mu| \sim M_{SUSY} \gg m_{EW}$) the SUSY QCD
corrections provide by far the dominant contribution to
the one-loop  MSSM correction. However, it is worth 
analyzing in more detail the SUSY EW corrections, which are known to provide extra
non-decoupling contributions. These nondecoupling SUSY EW effects
have been estimated in the literature just in the large $\tan\beta \gg
1$ regime and by means of an effective Lagrangian formalism,
which works in the zero external momentum approximation~\cite{cmwpheno,kolda,CarenaDavid}. In this approach, the
potentially large $\tan\beta$ enhanced SUSY corrections to the
Higgs boson-quark-quark Yukawa couplings are induced via the quark mass
corrections. In particular, the expression for the bottom quark Yukawa
coupling, at the one-loop level and in the large $\tan\beta$ limit, is~\cite{cmwpheno,CarenaDavid,EberlTodos}
\begin{equation}
h_b = \frac{m_b}{v} \tan\beta (1- \Delta m_b),
\label{eq:Yuk}
\end{equation}
where $m_b$ is the on-shell pole bottom quark mass, $v=174$ GeV, and the
$\tan\beta$ enhanced radiative corrections are encoded in 
\begin{equation}
\Delta m_b = \Delta m_b^{SQCD} + \Delta m_b^{SEW},
\end{equation}
where $\Delta m_b^{SQCD}$ and $\Delta m_b^{SEW}$ refer to the bottom quark mass corrections from the
SUSY QCD and SUSY EW sectors, respectively. $\Delta
m_b^{SQCD}$ is dominated by the bottom squark-gluino loops, and to leading order in
the strong coupling is~\cite{pierce,cmwpheno,CarenaDavid,EberlTodos}:
\begin{equation}
\Delta m_b^{SQCD} = \frac{2 \alpha_s}{3 \pi} M_{\tilde g} \mu
\tan\beta I(M_{\tilde b_1},M_{\tilde b_2},M_{\tilde g}).
\end{equation}
For sizable values of the soft trilinear coupling $A_t$,
$\Delta m_b^{SEW}$ is dominated by the top squark-chargino loops and more
precisely by the  top squark-charged higgsino contribution. By neglecting the
bino effects, which have been found to be numerically insignificant, the
SUSY EW mass correction to leading order in the top quark Yukawa coupling
and the electroweak gauge coupling is~\cite{pierce,cmwpheno,CarenaDavid},

\begin{eqnarray}
\Delta m_b^{SEW} &=& \frac{h_t^2}{16 \pi^2} A_t \mu
\tan\beta I(M_{\tilde t_1},M_{\tilde t_2},\mu) - \frac{g^2}{16 \pi^2} M_2 \mu
\tan\beta \left[ c_t^2 I(M_{\tilde t_1},M_2,\mu) \right. \nn \\ 
&&\left. + s_t^2 I(M_{\tilde t_2},M_2,\mu) + \frac{1}{2}c_b^2 I(M_{\tilde b_1},M_2,\mu)
+ \frac{1}{2}s_b^2 I(M_{\tilde b_2},M_2,\mu) \right], 
\end{eqnarray}
where $M_2$ is the SUSY soft breaking chargino mass parameter and $c_q
= \cos \theta_{\tilde q}$, $s_q=\sin\theta_{\tilde q}$, with
$\theta_{\tilde q}$ the $\tilde q$ mixing angle. The one-loop
integrals in $\Delta m_b^{SQCD}$ and $\Delta m_b^{SEW}$ are defined by,

\begin{equation}
I(m_1,m_2,m_3) = \frac{m_1^2 m_2^2 log\frac{m_1^2}{m_2^2} + 
m_2^2 m_3^2 log\frac{m_2^2}{m_3^2} + m_3^2 m_1^2 log\frac{m_3^2}{m_1^2}}
{(m_1^2-m_2^2)(m_2^2-m_3^2)(m_1^2-m_3^2)}.
\end{equation}

For the simplest assumption that is considered in this work
 of all SUSY soft breaking parameters
and the $\mu$ parameter being of comparable size, $M_{\tilde Q} \sim
M_{\tilde U} \sim M_{\tilde D} \sim M_{\tilde g} \sim M_{1,2} \sim
|A_{t,b}| \sim |\mu| \sim M_{SUSY} \gg m_{EW}$,
the loop integrals and mixing angles behave as,
\begin{eqnarray}
&& I(m_1,m_2,m_3) \sim  \frac{1}{2 M_{SUSY}^2} 
+ \mathcal{O}\left(\frac{m_{EW}}{M_{SUSY}^3} \right), \nn \\
&& c_q^2 \sim s_q^2 \sim \frac{1}{2} + 
\mathcal{O}\left(\frac{m_{EW}}{M_{SUSY}} \right),
\end{eqnarray}
and consequently the mass corrections behave as,
\begin{eqnarray}
\Delta m_b^{SQCD} &\sim& \frac{\alpha_s}{3 \pi} \frac{M_{\tilde
g}\mu}{M_{SUSY}^2} \tan\beta, \nn \\
\Delta m_b^{SEW} &\sim& \frac{h_t^2}{32 \pi^2} \frac{\mu
A_t}{M_{SUSY}^2}\tan\beta - \frac{3 g^2}{64 \pi^2} \frac{\mu
M_2}{M_{SUSY}^2} \tan\beta. 
\end{eqnarray}
First, we see here that the nondecoupling effects induced from the
previous expression for $\Delta m_b^{SQCD}$ 
on the bottom quark Yukawa coupling through eq.(\ref{eq:Yuk}) are exactly
the same as the ones extracted from the observables $O_{H}$ in eq.(\ref{eq.exprobs}) in the large $\tan\beta$
limit. Therefore both approaches coincide in this limit, but
just eq.(\ref{eq.exprobs}) gives the correct behavior for moderate and low
$\tan\beta$.
Second, we see that, even in the case of very large $\tan\beta$ values
and large $\mu$, $A_t$ and $M_2$, the size of the SUSY EW corrections remains
always well below the SUSY QCD corrections. The relative signs of
these contributions depend on our choice of the relative signs of
$M_{\tilde g}$, $\mu$, $A_t$, and $M_2$. By choosing a combination of
signs that maximizes the size of $\Delta m_b^{SEW}$ 
and for equally large SUSY parameters we find that $|\Delta
m_b^{SEW}|$ is less than
$ 50\%$ of $|\Delta m_b^{SQCD}|$.
Therefore, for this assumption on the size and signs of the SUSY parameters, 
we can conclude conservatively that the effect on the observables of
 eq.(\ref{eq.exprobs}) would
be reduced by at most $50\%$.

Since we are studying here the sensitivity of the observables in
eq.(\ref{eq.exprobs}) involving Higgs bosons and top quark decays to the SUSY QCD
nondecoupling corrections for all $\tan\beta$ values, a more precise analysis  of these SUSY EW
 nondecoupling effects would require the use of compact formulas
valid for all $\tan\beta$ values, which are not available so far in
the literature, and it is beyond the scope of this work. 
We will, however, take them into account conservatively in the final numerical analysis and
conclusion by considering 
this somewhat pessimistic case in which they conspire for a $50\%$ reduction of
the SUSY QCD signal.

\subsection{Predictions for $O^o$ in the minimal supersymmetric standard model}

Here we discuss the predictions for the leading contributions to the
observables, $O^o$, that do not include the SUSY particle contributions. 
Our purpose is to evaluate their uncertainties, and study whether or not they 
can mask the SUSY QCD corrections. Uncertainties below 1\% are neglected.

The leading contributions $O^o$ are computed here within the MSSM, and consist
of the tree level part plus the one-loop $\alpha_S$ corrections from 
standard QCD\footnote{This is to be consistent with the SUSY QCD 
$\alpha_S$ corrections.}. 
They are evaluated numerically with the FORTRAN program 
HDECAY~\cite{hdecay}, using the renormalization group 
approach of~\cite{higgsmasses} to obtain the MSSM Higgs boson masses. 
These standard QCD corrections contribute obviously just to the 
decays into quarks
and are known to be as large as $50\%$ (for a review, see~\cite{spiraQCD}). 
In order to take into account large contributions from higher orders, 
we have used
the running instead of the pole bottom quark mass. This resums the
leading logarithms and improves the convergence of the series.  To illustrate 
this, we show here the case of $A^o \to \bar b b$. We compare, 
in Table 1, the results of $\Gamma (A^o \to \bar b b)$
for $m_{A^o}=500$ GeV (normalized to the tree level width), 
computed at tree level, and at orders $\mathcal {O} (\alpha_S)$ and 
$\mathcal {O} (\alpha_S^2)$ in perturbation theory, using both the pole and 
running masses.
  
\begin{table}[ht]
\vskip 0.4cm
\begin{center}
\begin{tabular}{cccc}
\hline \hline
& tree-level & $\mathcal {O} (\alpha_S)$ & $\mathcal {O} (\alpha_S^2)$ \\
\hline
Pole mass & 1 & 0.523 & 0.369 \\ \hline
Running mass & 0.309 & 0.351 & 0.363 \\ \hline \hline
\end{tabular}
\label{tab.runmass}
\parbox{5in}{\caption [0]
{Comparison of QCD contributions to 
$\Gamma (A^o \to \bar b b)$, normalized to its tree level value, using 
the pole bottom quark mass and the running bottom quark mass, for $m_{A^o}=500$ GeV.}}
\end{center}
\end{table}

As can be seen from the results in this table, the convergence of the  
series is notably improved when the running mass is used. In addition, 
the error committed by ignoring $\mathcal {O} (\alpha_S^2)$ corrections is
reduced to values $\sim 3 \%$.

In the numerical evaluation of all the $O^o$ we will include, therefore, these
one-loop $\mathcal {O} (\alpha_S)$, QCD contributions and use the 
running bottom quark mass. Both the running bottom quark mass and $\alpha_S$ are evaluated 
at the corresponding particle decaying mass. 
We do not include, however, the standard electroweak 
radiative corrections 
from $W^{\pm}$, $Z$, $\gamma$, or extra Higgs bosons since they are known
to be below $\sim 1 \%$ in the MSSM (for a review, see \cite{spiraQCD} for the Higgs
bosons decay case and \cite{topEW} for the top quark decay case).

\begin{figure}
\begin{center}
\epsfig{file=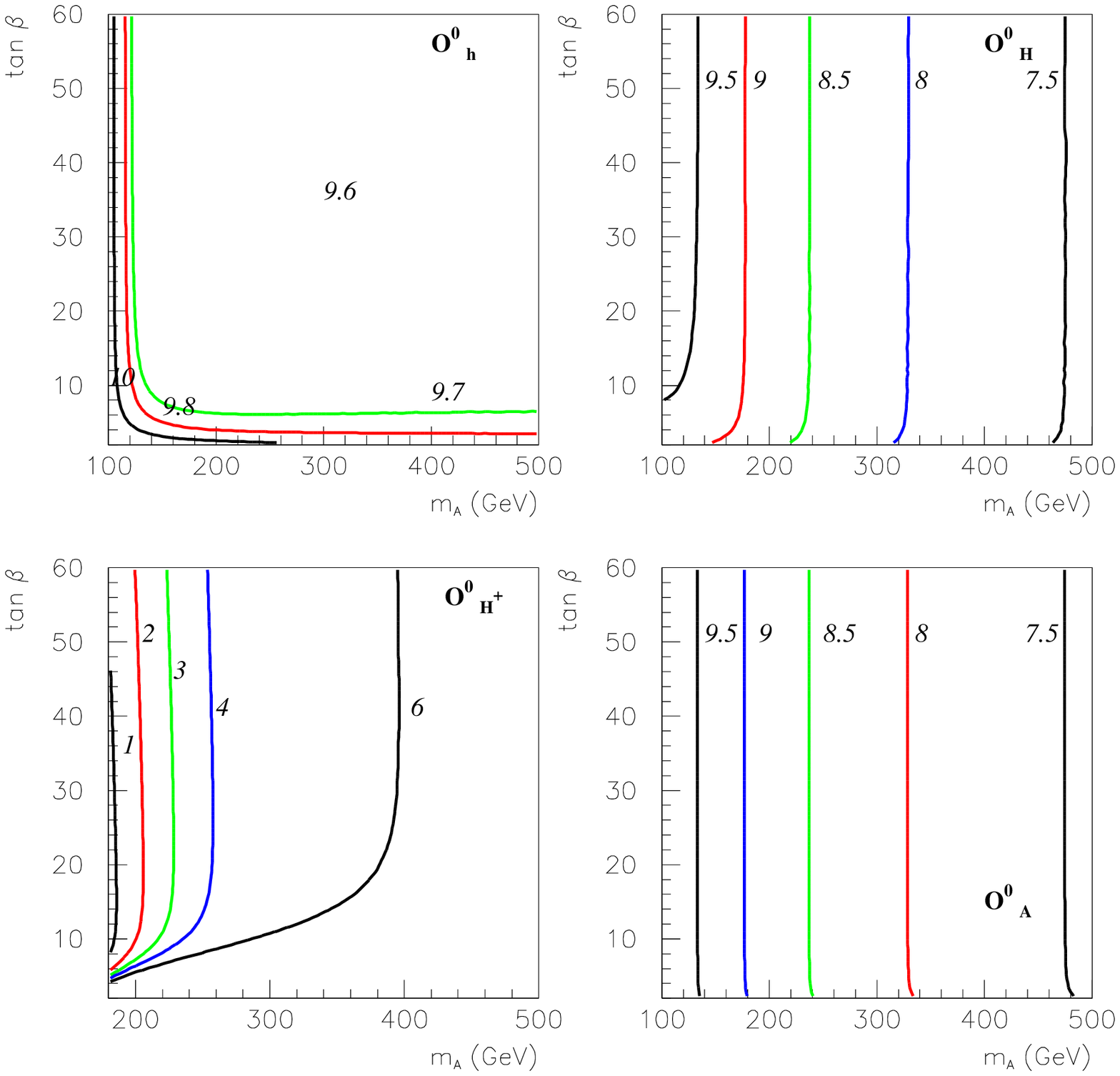,width=12cm}\\
\vskip 0.3cm
\epsfig{file=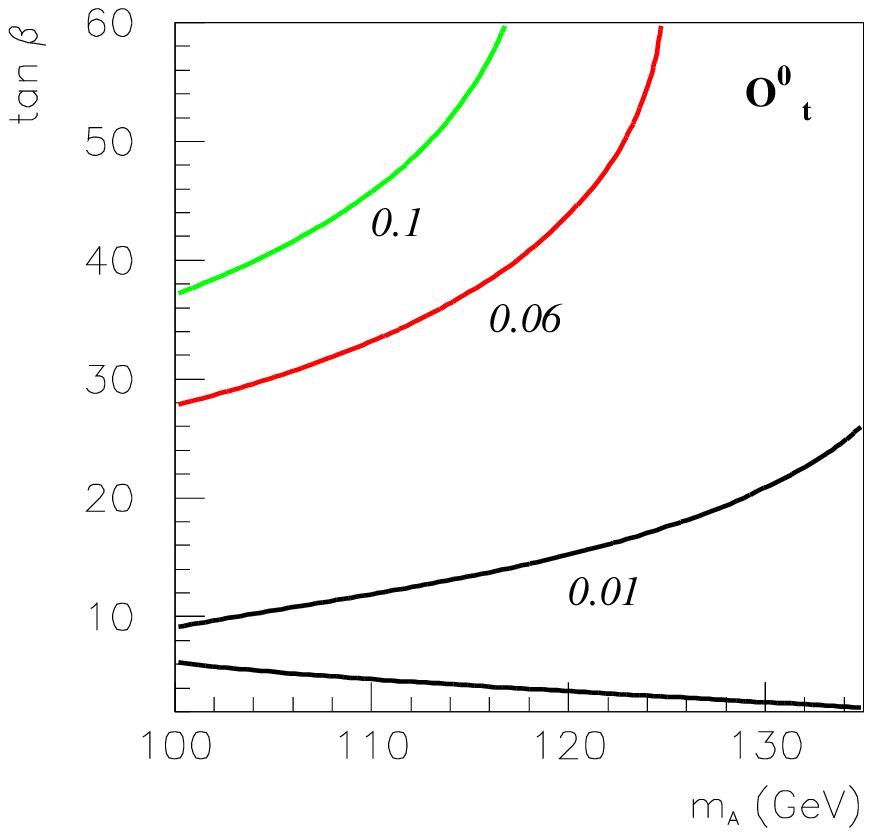,width=6cm}
\caption{\it  $O^o$ contour lines in the ($m_{A^o},\tan\beta$) plane predicted 
within the MSSM. The input parameters are defined in the text.}
\label{fig.sincor}
\end{center}
\end{figure}

The predictions for $O^o$ as a function of $m_{A^o}$ and $\tan\beta$ are shown 
in Fig.~\ref{fig.sincor}. The contour lines for $O^o$ in the 
($m_{A^o},\tan\beta$) plane
show clearly a very mild dependence on $\tan\beta$, for large $\tan\beta$
values, in all the Higgs boson channels.  In the neutral channels, this
comes exclusively from the dependence of the MSSM Higgs masses on $\tan\beta$.
The dependence on $m_{A^o}$ comes mainly from the running of the bottom quark mass,
which is evaluated at the corresponding Higgs boson mass. In the case of $h^o$, 
this dependence is frozen in the large $m_{A^o}$ region, where it behaves like the 
SM Higgs boson.
In the case of the $H^+$, the prediction grows significantly with $m_{A^o}$, as
$m_{H^+}$ separates from $m_t$.
Finally, for the top quark decay channel, the prediction changes rapidly with
$m_{A^o}$ and $\tan \beta$. 
The $\tan^2 \beta$ dependence can be seen 
in the large $\tan\beta$ region. Moving to the right part of the plot, the 
size of the predictions decreases as $m_{H^+}$ approaches $m_t$.

We next analyze the theoretical uncertainties in the evaluation 
of the ratios in 
eq.(\ref{eq.interobsH}) and eq.(\ref{eq.interobst}) coming from the experimental errors in the values of 
the SM parameters 
involved in their determination, at one-loop and order $\alpha_S$. We have 
found that the only significant errors are the ones coming from 
errors in the bottom and top quark masses and in the strong coupling constant 
$\alpha_S$. We have used the results from~\cite{pdg}:

\[ m^{\rm pole}_b = (4.6 \pm 0.2)\, {\rm GeV}, 
\quad m_t = (174.3 \pm 5.1) \, {\rm GeV}, 
\quad \alpha_S (M_Z^2) = 0.118 \pm 0.002 \nn\]

The uncertainty coming from these experimental errors is
$\delta O=\frac{O'-O}{O}$, where $O$ is the observable evaluated 
taking central values for $m_b$, $m_t$, and $\alpha_S$, and $O'$ is the 
observable evaluated with one parameter reaching one extreme value. 
In Table 2 the different uncertainties are
summarized for three different values of $m_{A^o}$, in the region
$\tan\beta = 10 - 60$. We use $\delta O_n \equiv \delta O_{h^o},\ 
\delta O_{H^o},\ \delta O_{A^o}$, and $\delta O_c \equiv \delta O_t$ for  
$m_{A^o} = 100$ GeV, while $\delta O_c \equiv \delta O_{H^+}$ for 
$m_{A^o} = 250$ and $500$ GeV.

Notice that the corrections are approximately constant as a function 
of $\tan \beta$.
The uncertainty related to the 
error in $m_b$ is the dominant one, and of the same size for all the 
observables. Finally, the uncertainty related to $m_t$ is significant 
for the charged Higgs boson and top quark cases, only in the vicinity of the corresponding kinematical
thresholds.
The total uncertainty in the sixth column has been evaluated as the sum
in quadrature of the numbers in the previous three columns and the error 
coming from neglecting the ${\mathcal O}(\alpha^2_S)$ corrections.

\begin{table}[ht]
\begin{center}
\begin{tabular}{cccccc}
\hline
\hline
 $\delta$ &$m_{A^o}$ & $\alpha_s$ & $m_b$ & $m_t$ & Total \\ \hline
 & 100 GeV & $2\%$ & $11\%$ & $<1\%$ & $12\%$ \\ 
 $\delta O_n$ & 250 GeV& $3\%$ & $10\%$ & $<1\%$ & $12\%$ \\ 
  & 500 GeV & $3\%$ & $11\%$ & $<1\%$ & $12\%$ \\ \hline
& 100 GeV & $1\%$ & $8\%$ & $6\%$ & $11\%$ \\ 
 $\delta O_c$ & 250 GeV& $1\%$ & $10\%$ & $7\%$ & $13\%$ \\ 
  & 500 GeV & $3\%$ & $11\%$ & $<1\%$ & $12\%$ \\ \hline \hline
\end{tabular}
\parbox{5in}{\caption [0]
{\label{sumtable} Uncertainties in the ratios induced by the experimental 
 errors in $m_t$, $m_b$, and $\alpha_S$ for three values of $m_{A^o}$, 
in the region $\tan\beta = 10 - 60$.  
We use $\delta O_n \equiv \delta O_{h^o}, 
\delta O_{H^o}, \delta O_{A^o}$ and $\delta O_c \equiv \delta O_t$ for  
$m_{A^o} = 100$ GeV, 
and $\delta O_c \equiv \delta O_{H^+}$ for $m_{A^o} = 250, 500$ GeV. 
The total uncertainty has been evaluated in
quadrature and includes the error coming from neglecting the
${\mathcal O}(\alpha^2_S)$ corrections.}}
\end{center}
\end{table}

In summary, the MSSM predictions for the previous Higgs boson observables are 
known up to an uncertainty of the order of 13\%. Any extra correction must 
be larger than that to be visible. 
 
\subsection{Comparing a type II two higgs doublet model and the minimal supersymmetric standard model}

 Here we investigate the similarities and differences of 
 the set of observables defined in Section 3.1, in a general 2HDMII 
 as compared to the MSSM.
 For a given Higgs sector mass pattern that is compatible with 
 the MSSM, the values of the observables
 at the tree level, and after including the standard QCD corrections, coincide
 with the corresponding ones in the MSSM.
 Note that the mixing angle in the neutral 
 sector, $\alpha$, is a derived quantity in the MSSM, whereas in the 2HDMII 
 it is an independent parameter (in principle $-\pi/2\leq\alpha\leq\pi/2$,
 but actually the range in $\alpha$ is further restricted if the Higgs
 potential is required to be bounded from below) and, hence, 
 its value can be different in the two models.
 However, since the branching 
 ratios in the numerator and denominator 
 of our observables $O$ have the same dependence on $\alpha$, at the
 tree level the
 $O^{\rm tree}$ themselves are $\alpha$ independent 
 (see eq.(\ref{eq.treeobserv})). When the standard QCD 
 corrections are included, this $\alpha$ independence is still maintained. 
 On the other hand, the 
 standard electroweak corrections from $W^{\pm}$, $Z$, $\gamma$ 
 are also similar in both models and, as   
 already said,  they have been estimated to be below $1\%$ and 
 can be ignored here. Therefore, the only differences, apart from the SUSY
 corrections, could
 come from the Higgs sector radiative corrections. As mentioned before,
 these are negligible in the MSSM but, in principle, could be 
 larger in the 2HDMII.
 The reason is that these corrections involve the Higgs boson self-couplings which
 in the 2HDMII are not restricted by SUSY and, in principle, can be large.
 We have estimated these potentially different contributions from the Higgs 
 sector to the one-loop level and for a Higgs boson mass pattern compatible with 
 the MSSM, and they turn out to be very small, certainly well below the 
 common ones discussed in the previous section.
 
\begin{figure}[h]
\begin{center}
\epsfig{file=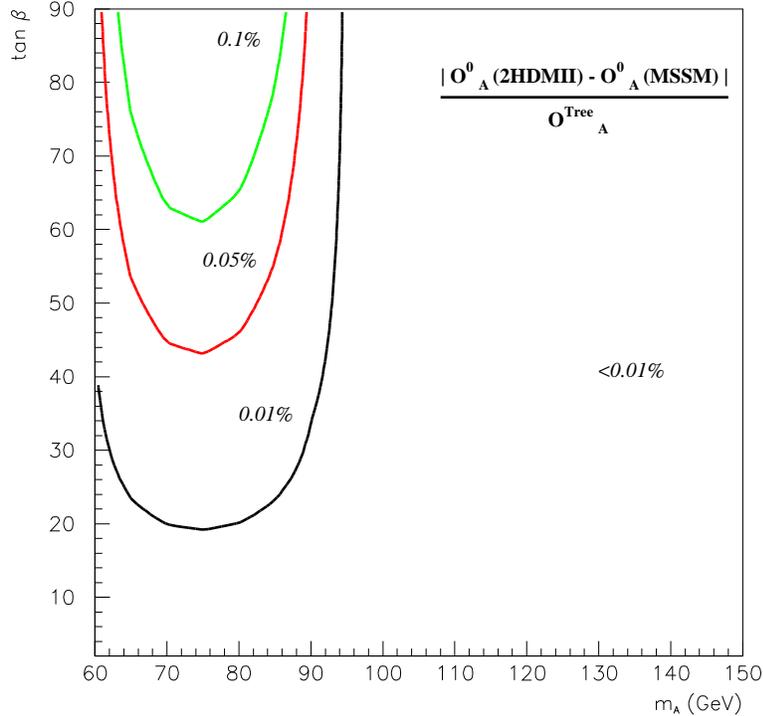,width=10cm}
\caption{\it  The maximum absolute value of the difference 
 between the MSSM and 2HDMII Higgs sector corrections for $O^o_{A^o}$
 (after scanning the allowed range in $\alpha$),
 relative to the tree-level prediction,  
 $O^{\rm tree}_{A^o}$. Note that the area 
 $m_{A^o}< 84.1$ GeV (if $\tan\beta >1$) is already 
 experimentally excluded at $95\%$ CL~\cite{pdg}.}
\label{fig.diff}
\end{center}
\end{figure}

 In order to illustrate this we discuss the case of $A^o$.  We show 
 in Fig.\ref{fig.diff} the maximum absolute value of the difference 
 between the MSSM and 2HDMII Higgs sector corrections for $O^o_{A^o}$
 (after scanning the allowed range in $\alpha$),
 relative to the tree-level prediction  
 $O^{\rm tree}_{A^o}$. More specifically, these differences come exclusively 
 from the $\alpha$-dependent one-loop diagrams, and from the 
 triangular one-loop diagrams involving the Higgs boson self-couplings. 
 The contour lines separate the regions in the ($m_{A^o},\tan\beta$) plane 
 where the maximum difference between the MSSM and 2HDMII 
 is larger than 0.1\%, between 0.05\% and 0.1\%, between 
 0.01\% and 0.05\%, and below 0.01\%. In the region of 
 interest, $m_{A^o}>84.1$ GeV~\cite{pdg}, this difference is always less than 0.1\%.
   
 In summary, as announced, the basic difference between the MSSM and a general
 2HDMII model, with respect to our observables, is that the large SUSY particle 
 loop corrections do not exist in the second case. In the next section we
 discuss these contributions.
            
\section{Results and discussion}
\begin{figure}[ht]
\begin{center}
\epsfig{file=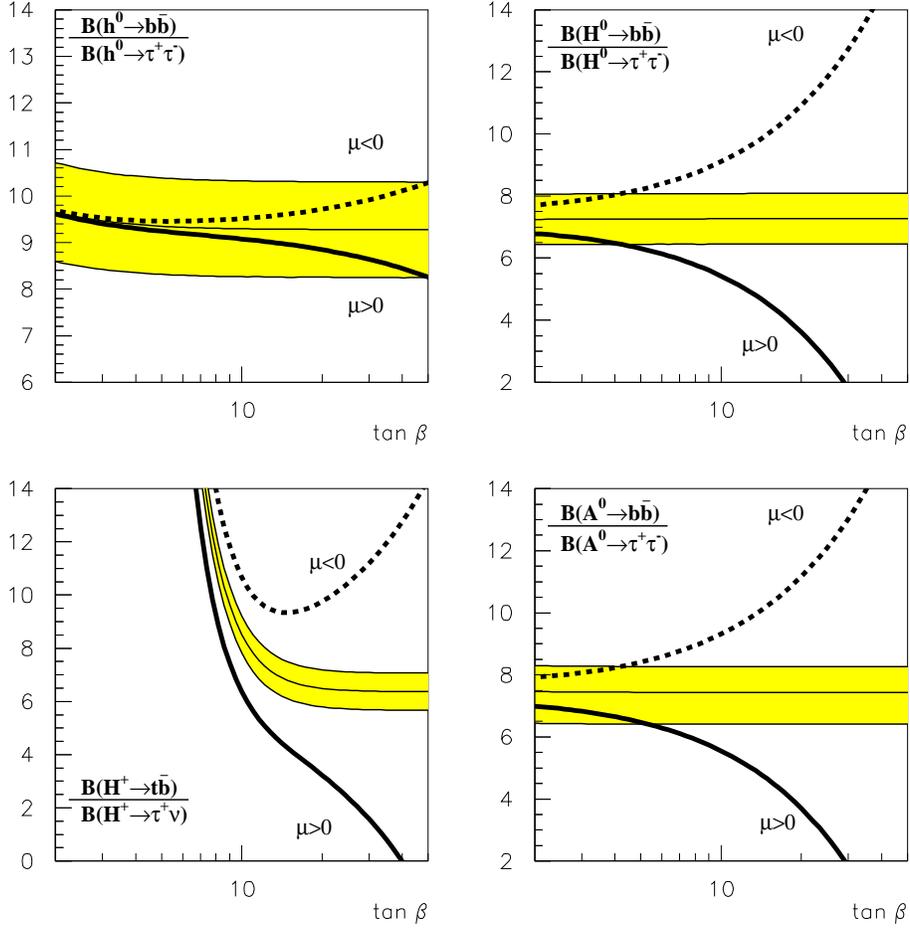,width=12cm} 
\caption{\it Predictions for $O_{h^o}$, $O_{H^o}$, $O_{H^+}$, and
$O_{A^o}$ as a function of
 $\tan\beta$, for $m_{A^o}=500$ GeV. The central lines are the corresponding
 predictions for $O^o$. The shaded bands cover the theoretical
 uncertainties estimated in Sect. 3.3. The bold lines represent the SUSY QCD
corrected predictions for $M_{\rm SUSY}=M_{\tilde g}=|\mu|$.}
\label{fig.sitges500}
\end{center}
\end{figure}

\begin{figure}[ht]
\begin{center}
\epsfig{file=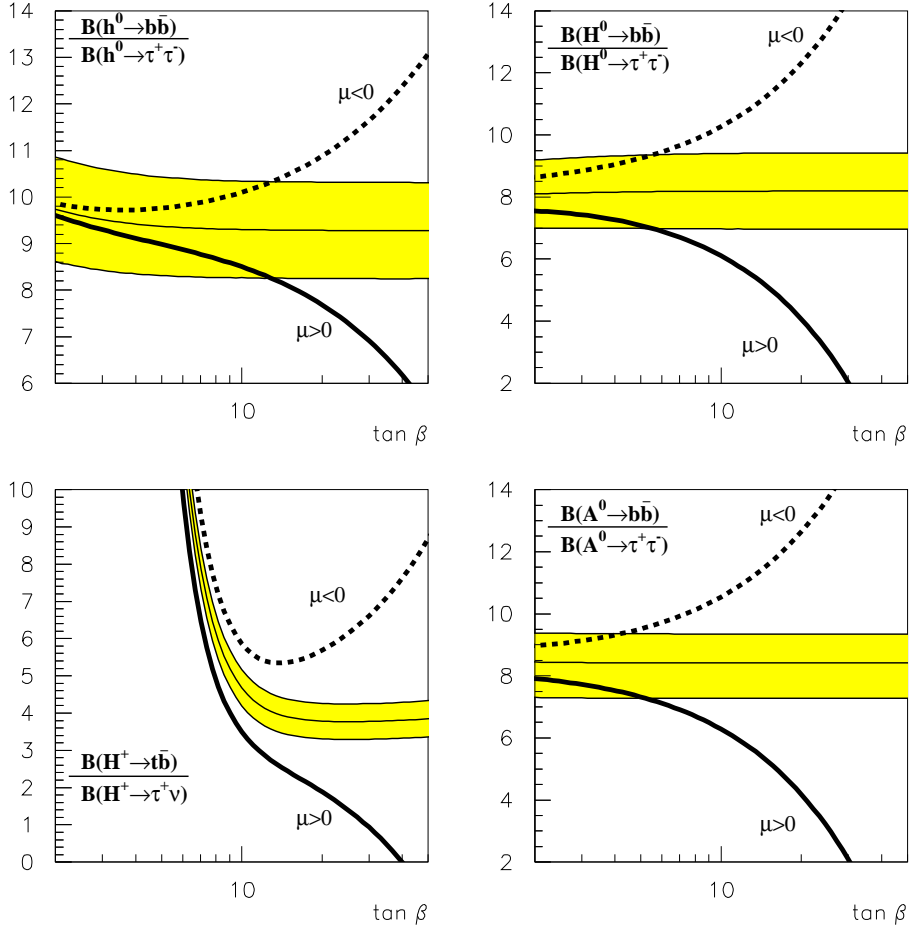,width=12cm} 
\caption{\it Same as in Fig.~\ref{fig.sitges500}, but for $m_{A^o}=250$ GeV.}
\label{fig.sitges250}
\end{center}
\end{figure}

\begin{figure}[ht]
\begin{center}
\epsfig{file=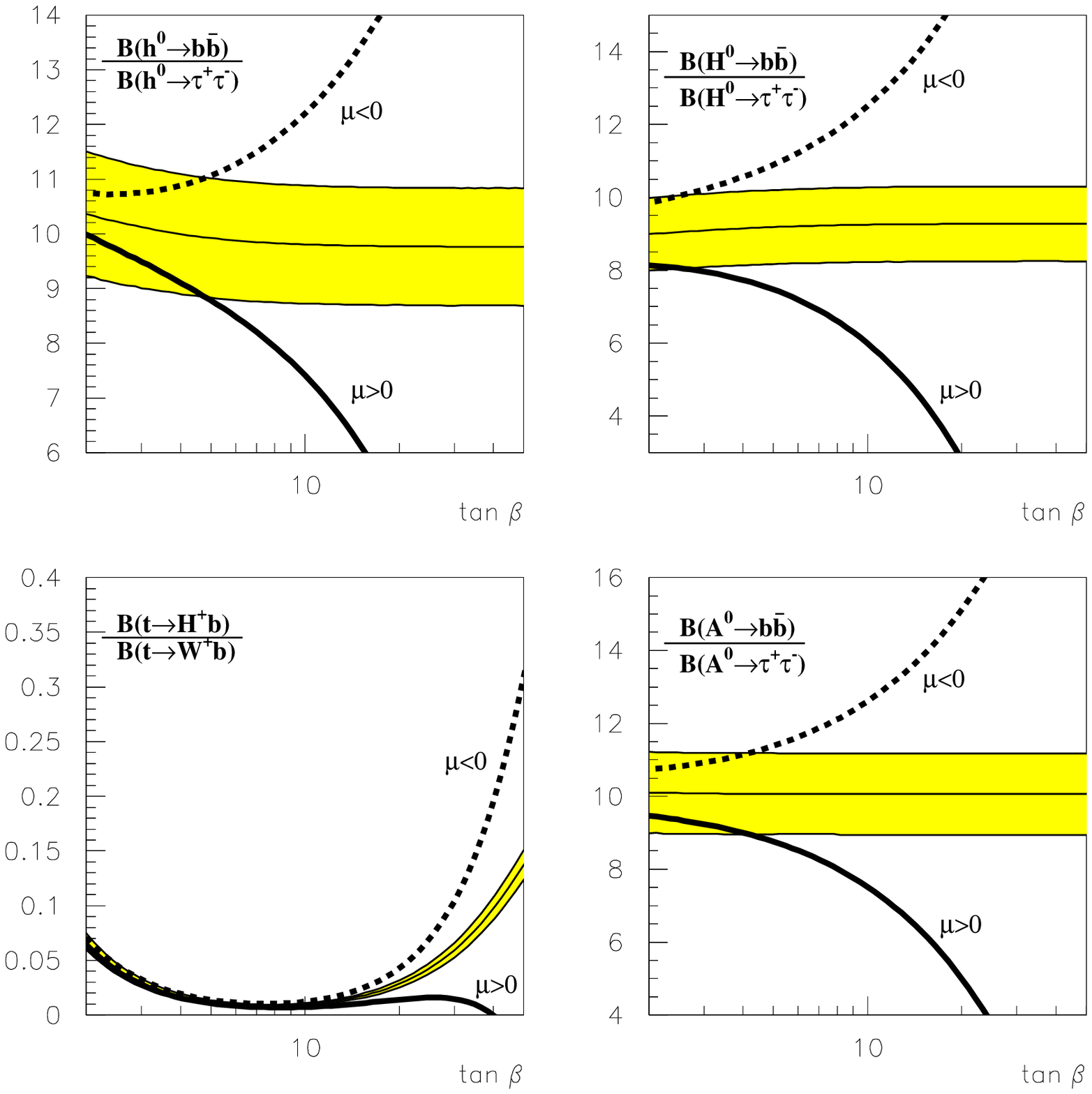,width=12cm} 
\caption{\it Same as in Fig.~\ref{fig.sitges500}, but for $m_{A^o}=100$ GeV. 
The prediction for $O_t$ appears in the lower left plot instead of 
$O_{H^+}$, since in this case $m_t > m_{H^+}+m_b$.}
\label{fig.sitges100}
\end{center}
\end{figure}

In this section we present  the numerical results for the SUSY QCD
corrections to the set of observables $O_{h^o}$, $O_{H^o}$, $O_{A^o}$, 
$O_{H^+}$, and $O_{t}$.  The relevant parameters in this discussion are
$\tan\beta$, $m_{A^o}$, $\mu$, and $M_{\rm SUSY}$ 
($M_{\rm SUSY}\sim M_{\tilde q} \sim
M_{\tilde g}$). We first discuss the 
behavior with $\tan\beta$. 
In order to focus on this $\tan\beta$ dependence, we consider  
the simplest choice for the SUSY mass parameters, that is, 
$M_{\rm SUSY}=M_{\tilde g}=|\mu|$. We show in 
Figs.~\ref{fig.sitges500}-\ref{fig.sitges100} the results for 
three $m_{A^o}$ values in the large, medium, and low $m_{A^o}$ regions: 
$m_{A^o}=500$, $250$, and $100\,{\rm GeV}$, respectively. The central
lines in these figures 
follow the predictions for the observables without the SUSY QCD contribution, 
namely, $O^o$. The corresponding total theoretical uncertainties,
discussed in Section 3.2, are shown as shaded
bands around the central values. The bold lines represent the SUSY QCD
corrected predictions for $\mu > 0$ (solid) and $\mu < 0$ (dashed).
The observable $O_t$ appears in the lower left plot of Fig.~\ref{fig.sitges100} 
replacing $O_{H^+}$, since in this case $m_t > m_{H^+}+m_b$.
 
One can see from the figures that, first,
the predictions for all the observables separate from the central values
as $\tan\beta$ grows. The sign of the SUSY QCD corrections is  positive 
for $\mu<0$ and negative for $\mu>0$. The central values and their 
theoretical uncertainties 
are rather insensitive to $\tan\beta$ in the $H^o$ and $A^o$ channels. 
For the $h^o$, there is a very slight dependence at low $\tan\beta$, while 
for the $H^+$ this dependence is very strong in the low $\tan\beta$ region 
and it softens for $\tan \beta>15$, as expected from 
eq.~(3).
In any case, the predictions for all the Higgs boson observables without the 
SUSY QCD contributions are $\tan\beta$ insensitive in the large 
$\tan\beta$ region. This is very different for the top quark observable:
it depends strongly on $\tan \beta$ in all the parameter space.

Concerning the size of the SUSY QCD corrections, we see that there
is always a $\tan\beta$ region where these are larger than the theoretical
uncertainties. For $m_{A^o}=500$ GeV and $m_{A^o}=250$ GeV, the 
observables $O_{A^o}$ and $O_{H^o}$ behave similarly and
the predictions are outside the shaded band 
for $\tan\beta>5$. For $m_{A^o}=100$ GeV, the crossing still 
happens at $\tan\beta>5$ for $A^o$, whereas for $H^o$ the 
bold lines lie outside the error band in the whole region
$2<\tan\beta<50$.
  
In the $h^o$ case the situation is qualitatively different.
For $m_{A^o}=500$ GeV, the SUSY QCD correction is below the 
theoretical uncertainty for all values in the region $2<\tan\beta<50$. 
This is a manifestation of the decoupling of this correction for 
large $m_{A^o}$ values, as commented in Section 3.1. 
For smaller values of $m_{A^o}$, such as $m_{A^o}= 250, 100$ GeV, the 
decoupling has not effectively operated yet and the SUSY QCD corrections 
are sizable for large enough $\tan\beta$. In particular, for 
$m_{A^o}=250\ (100)$ GeV, they are larger than the theoretical error band for   
$\tan\beta>15\ (5$).
 
Regarding the charged Higgs boson case, two different situations must be considered. For
$m_{A^o}=100$ GeV, where the decay into a top and a bottom quark is not kinematically
allowed,  the observable $O_t$ is considered. Otherwise, for $m_{A^o}=250$ GeV and 
$m_{A^o}=500$ GeV, the relevant observable is $O_{H^+}$.  
As can be seen in Fig. \ref{fig.sitges100}, the SUSY QCD corrections in 
$O_t$ for $m_{A^o}=100$ GeV are above the theoretical uncertainty for $\tan\beta>15$. However, 
as pointed out in Sect.~2, the central value prediction also depends strongly on 
$\tan\beta$ 
(and $m_{A^o}$) and it will be difficult to identify the effect of the SUSY QCD corrections above the
additional uncertainty related to the experimental errors on the measurement of $\tan\beta$
and $m_{A^o}$.
For $O_{H^+}$ and $m_{A^o}=500, 250$ GeV
both requirements, the correction being larger than the theoretical error and the 
leading contribution being insensitive to $\tan \beta$, are satisfied for values larger 
than about 15.  
  
\begin{figure}
\begin{center}
\epsfig{file=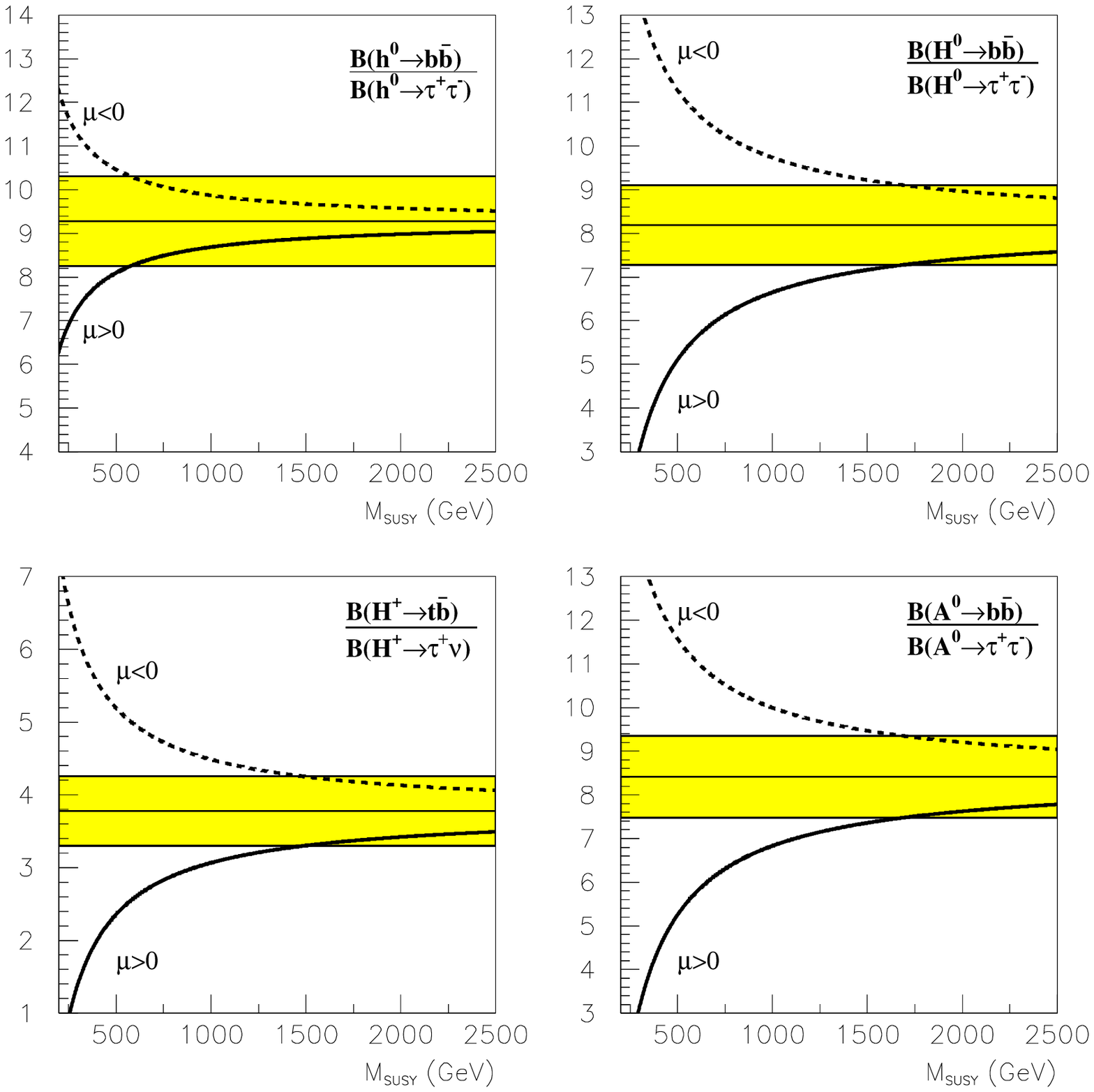,width=12cm}\\
\vskip 0.3cm
\epsfig{file=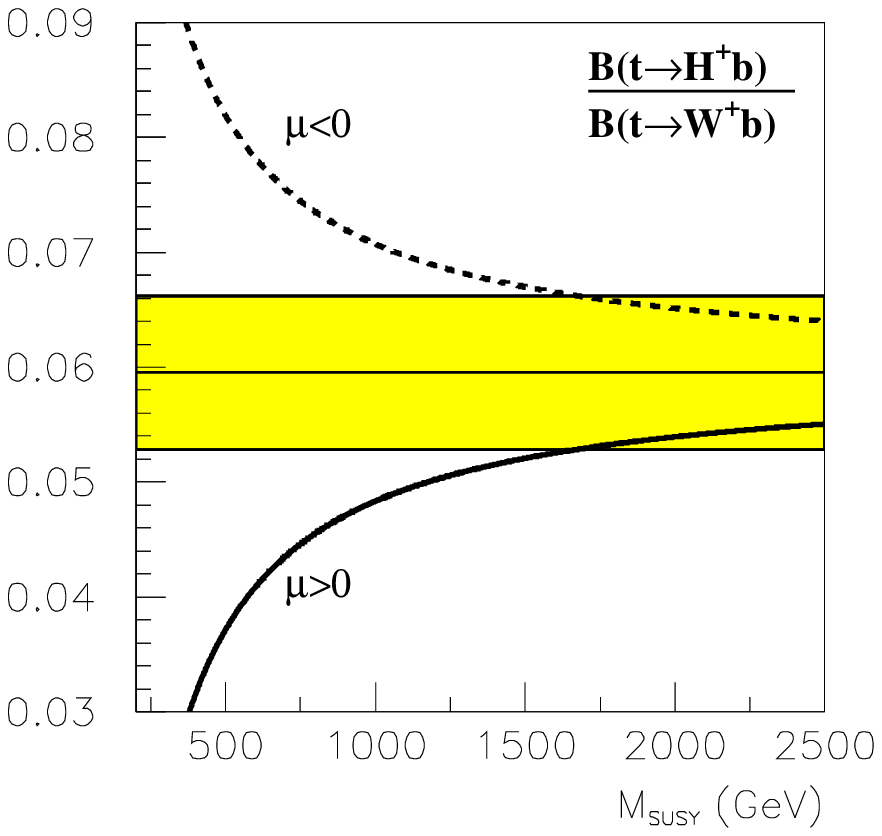,width=6cm}
\caption{\it Predictions for $O_{h^o}$, $O_{H^o}$, $O_{H^+}$, $O_{A^o}$, and $O_t$
 as a function 
 of $M_{\rm SUSY}=M_{\tilde g}$, for $|\mu|= 250$ GeV and $\tan \beta = 30$. 
 $m_{A^o}$ is fixed to 250 GeV for the observables $O_{h^o}$, $O_{H^o}$,
 $O_{H^+}$, and $O_{A^o}$. For $O_t$, $m_{A^o}=100$ GeV.}
\label{fig.sitgesMsusy}
\end{center}
\end{figure}

We next discuss the $M_{\rm SUSY}$ dependence. For this purpose, we fix the 
$|\mu|$ value to 250 GeV and choose $M_{\tilde g}=M_{\rm SUSY}$. 
Figure~\ref{fig.sitgesMsusy} shows
the predictions for the observables as a function of $M_{\rm SUSY}$ for 
$m_{A^o}=250$ GeV, $\tan\beta=30$, and both signs of the $\mu$ parameter. 
Since, in this case,  the SUSY QCD corrections vary as 
$1/M_{\rm SUSY}$, their size will be below the theoretical uncertainty
for large enough $M_{\rm SUSY}$. This occurs in Fig.~\ref{fig.sitgesMsusy}  
for $O_{H^o}$, $O_{A^o}$, $O_{H^+}$, and $O_t$ at about 1500 GeV.
For  $O_{h^o}$ it is at about 500 GeV. Therefore, except in the light Higgs boson
case, we expect the observables to be sensitive to indirect signals of supersymmetry
via these SUSY QCD corrections up to quite large values of $M_{\rm SUSY}$.    

\begin{figure}
\begin{center}
\epsfig{file=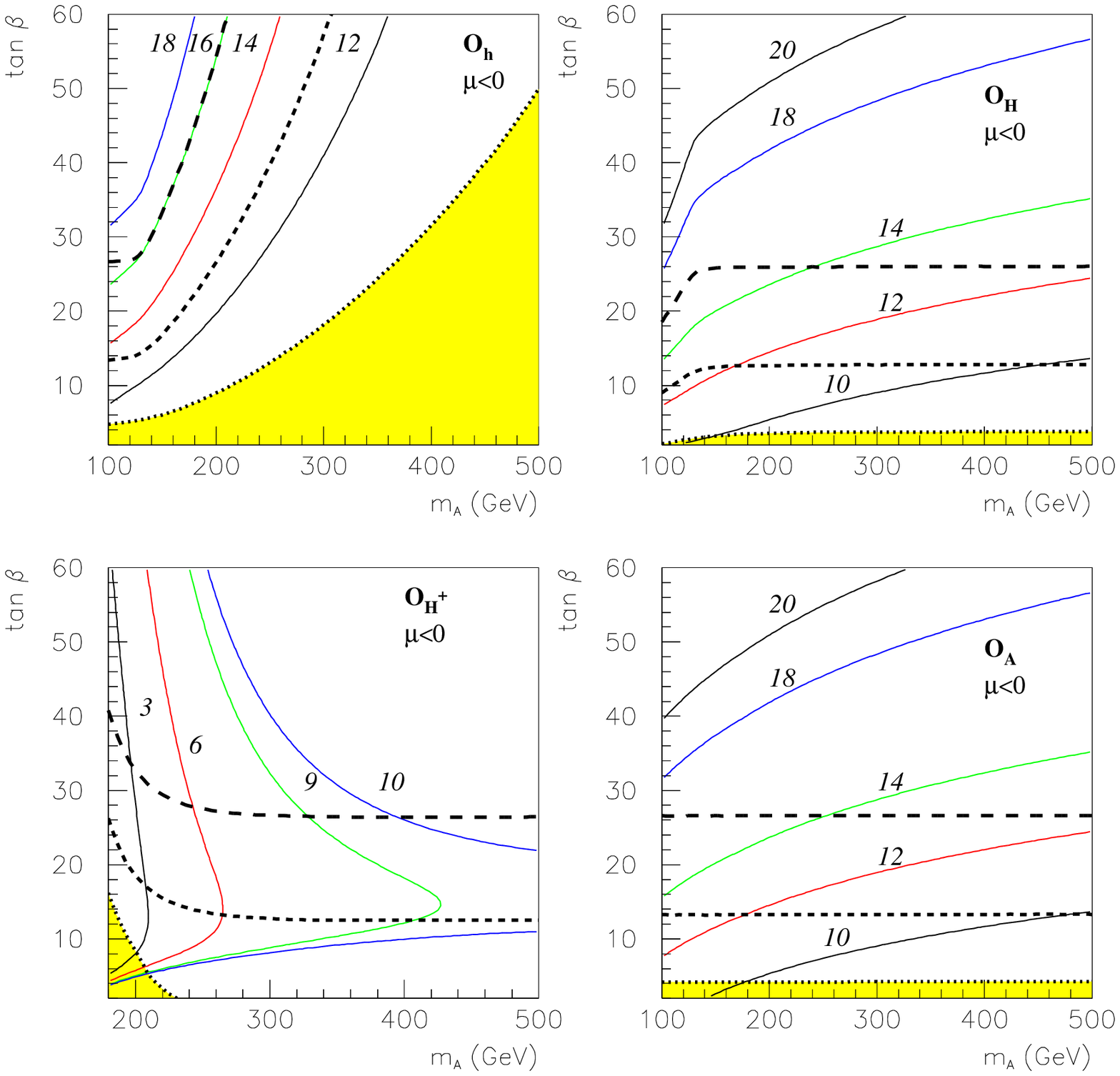,width=12cm}\\
\vskip 0.3cm
\epsfig{file=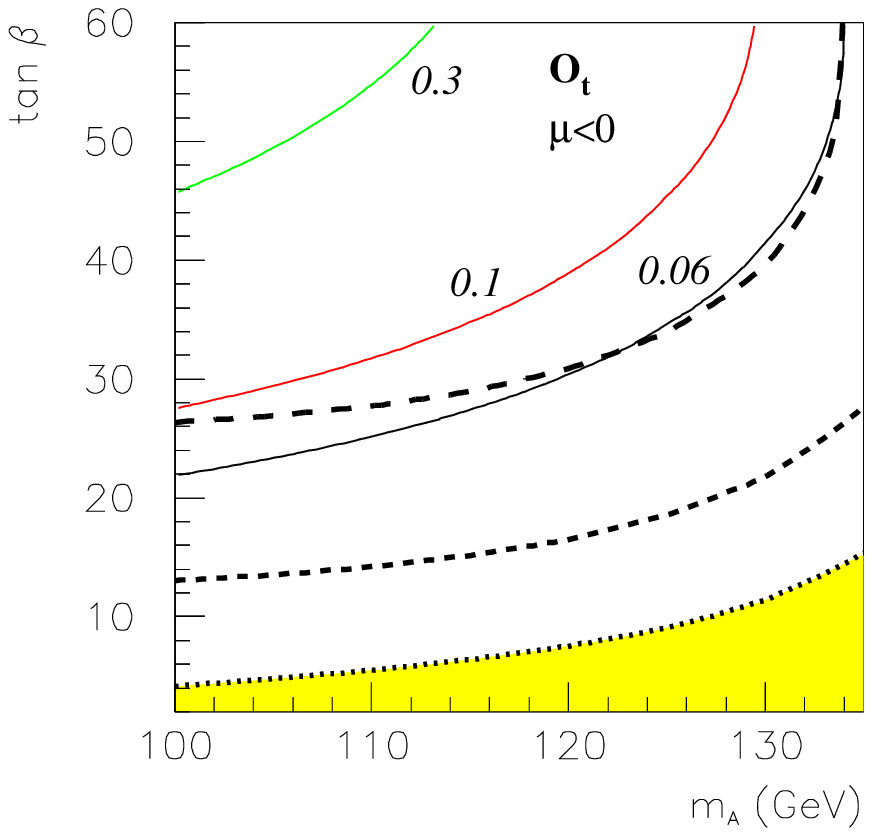,width=6cm}
\caption{\it Predictions for the observables including the SUSY QCD corrections
for $M_{\rm SUSY}=M_{\tilde g}=|\mu|$ and $\mu<0$. 
The solid contour lines follow the points in the $(m_{A^o},\tan\beta)$
plane with constant
value of $O$. The shaded area represents the region where the corrections are smaller
than the theoretical uncertainty. The long (short) dashed lines join the points where an 
experimental resolution of 50\% (20\%) is required to achieve a meaningful 
measurement.}
\label{fig.concorrneg}
\end{center}
\end{figure}

\begin{figure}
\begin{center}
\epsfig{file=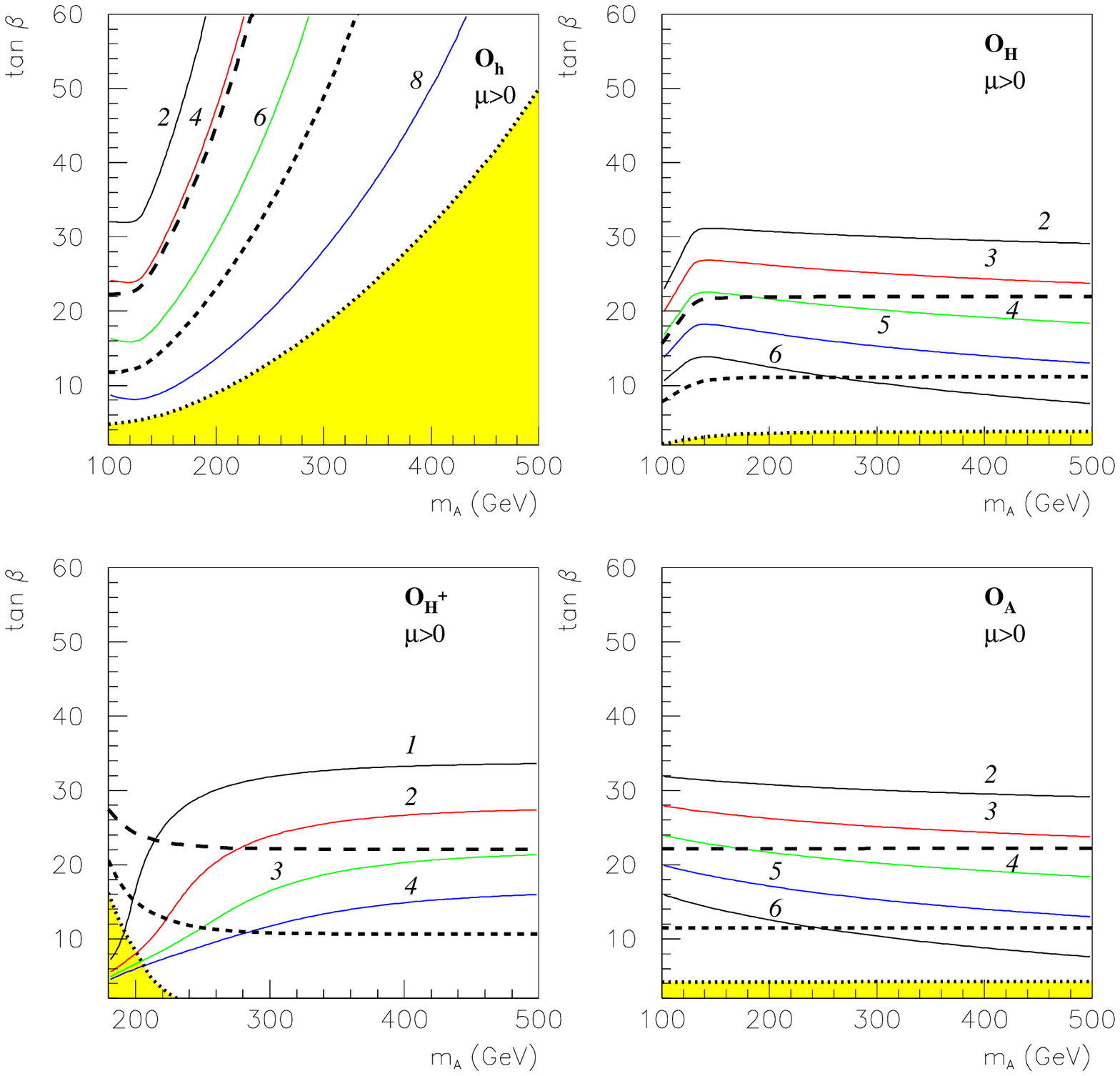,width=12cm}\\
\vskip 0.3cm 
\epsfig{file=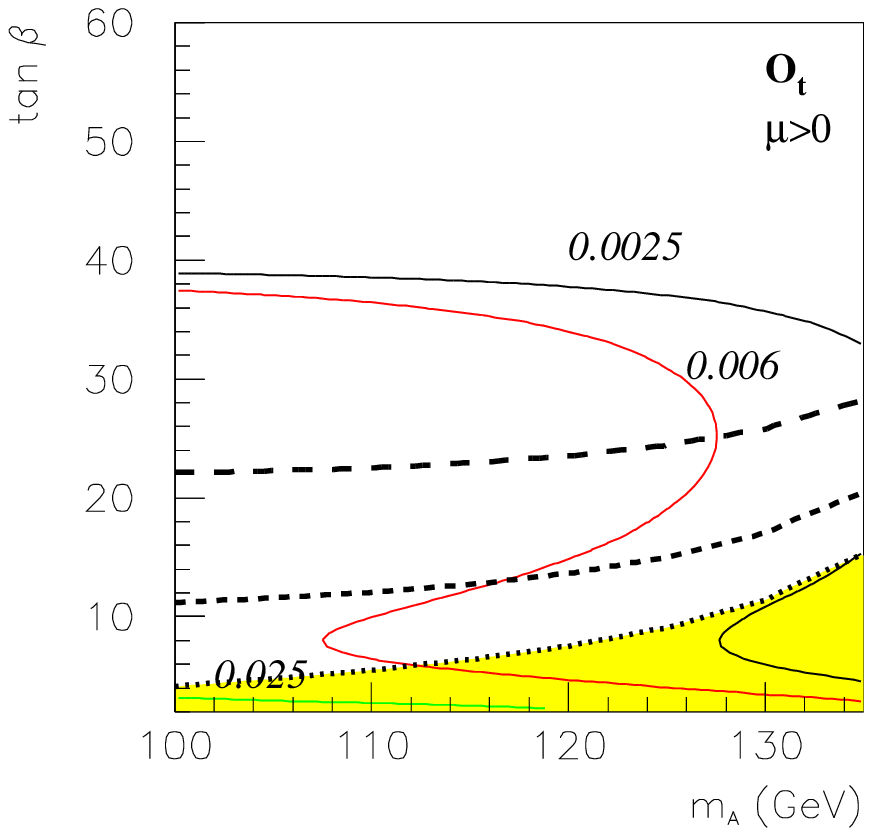,width=6cm}\\ 
\caption{\it Same as in Fig.\ref{fig.concorrneg}, but for $\mu>0$.}
\label{fig.concorr}
\end{center}
\end{figure}

\begin{figure}
\begin{center}
\epsfig{file=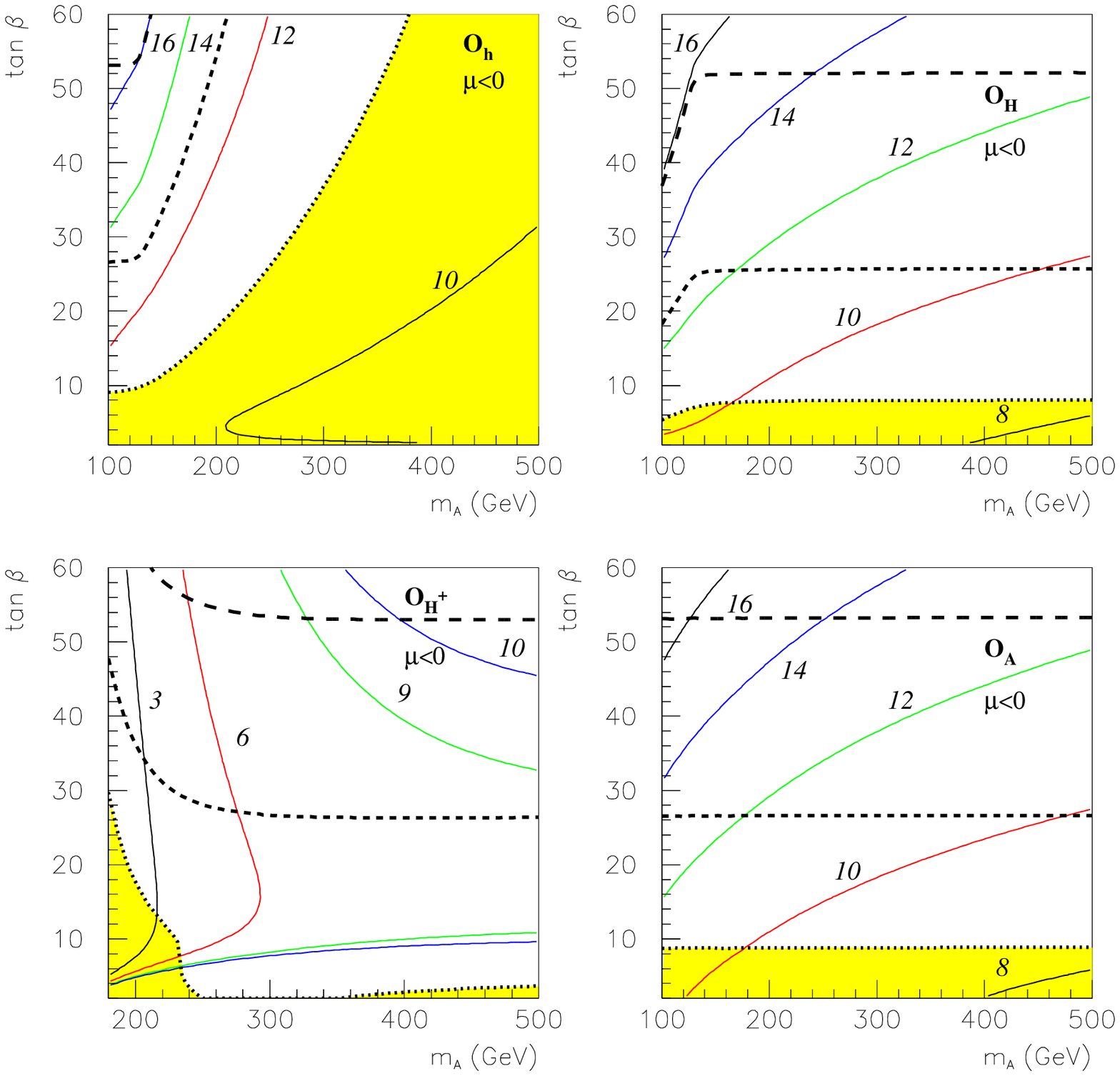,width=12cm}\\
\vskip 0.3cm
\epsfig{file=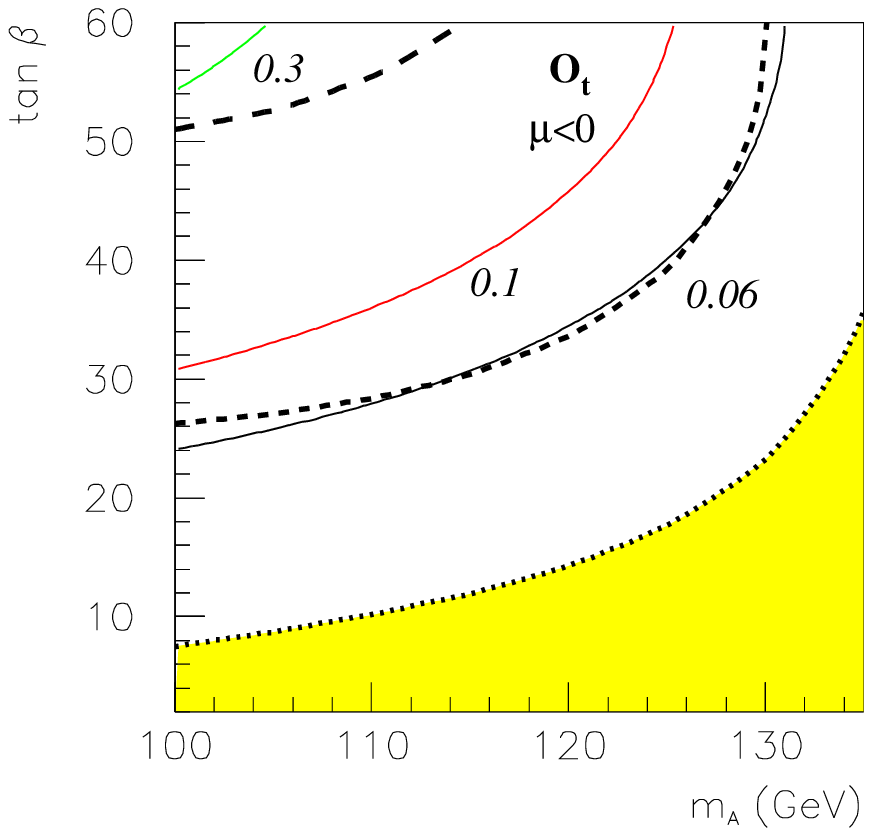,width=6cm}
\caption{\it Predictions for the observables with the SUSY QCD corrections
 reduced by $50 \%$ in order to simulate the effect of the SUSY EW contributions in the conservative scenario discussed in the text. The rest of the inputs and specifications are as in Fig.~\ref{fig.concorrneg}.
}
\label{fig.concorrnegew}
\end{center}
\end{figure}

\begin{figure}
\begin{center}
\epsfig{file=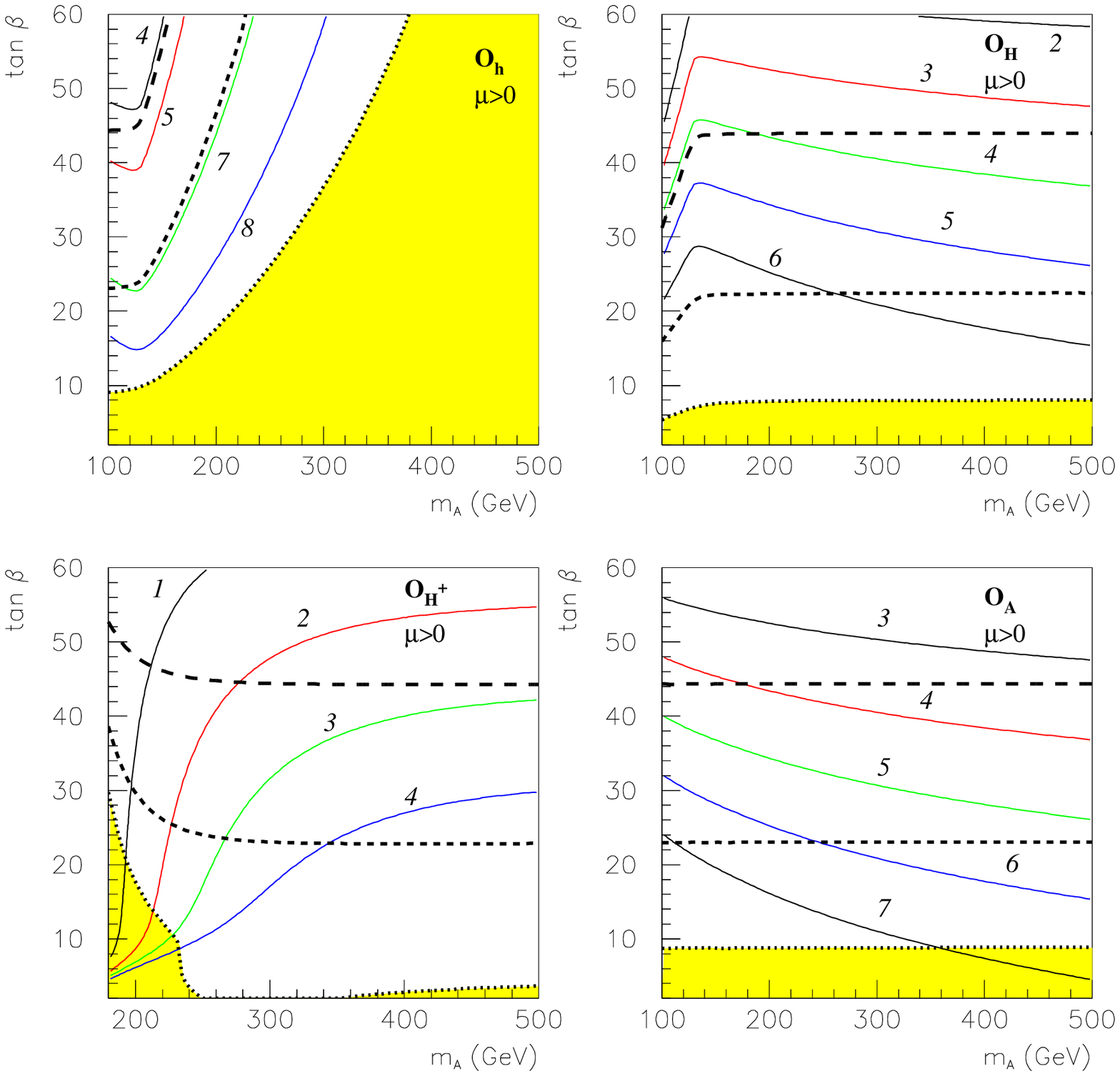,width=12cm}\\
\vskip 0.3cm 
\epsfig{file=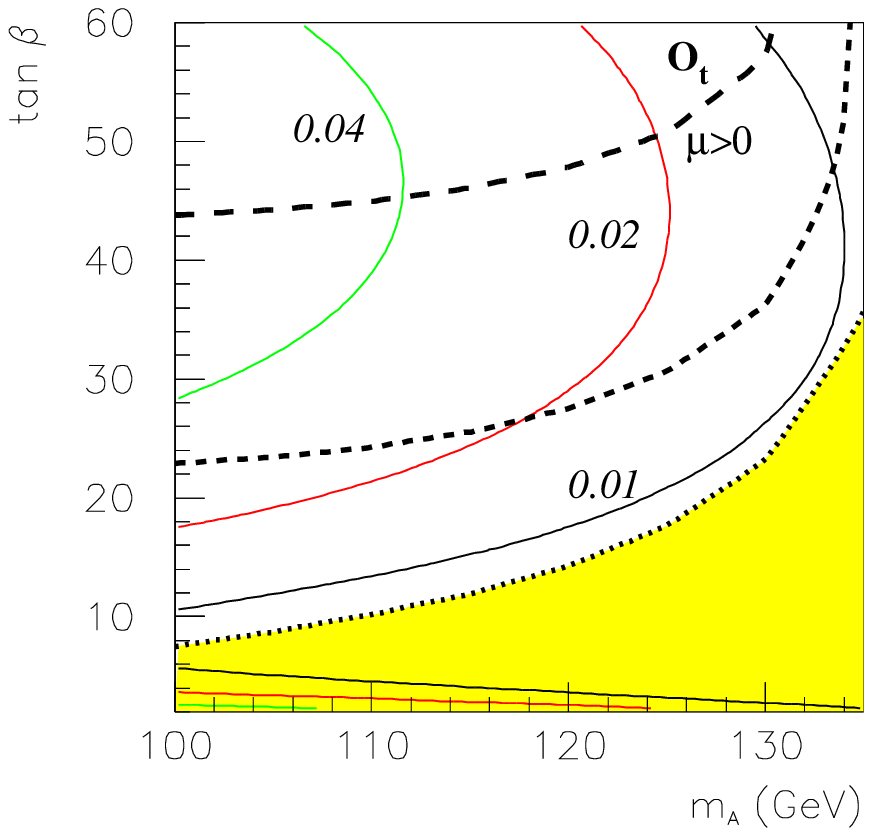,width=6cm}\\ 
\caption{\it Same as in Fig.\ref{fig.concorrnegew}, but for $\mu>0$.}
\label{fig.concorrew}
\end{center}
\end{figure}

We finally turn to the question of the experimental resolution required 
to evidence the above effects, after a Higgs boson signal is found at a given position in
the $(m_{A^o},\tan\beta)$ parameter plane. The predictions for the observables including 
the SUSY QCD corrections for $M_{\rm SUSY}=M_{\tilde g}=|\mu|$ are shown
in Fig.~\ref{fig.concorrneg}  for $\mu<0$ and in Fig.~\ref{fig.concorr} for $\mu>0$. 
Comparing these figures with Fig.~\ref{fig.sincor}, the patterns of the $O$ (solid) 
contour lines change noticeably with respect to the ones for $O^o$. The differences
are mainly because of both the large size of the corrections and the different behavior 
with $\tan\beta$. 

The shaded zone in each plot represents the region where the corrections (if existing) 
are totally hidden inside the theoretical uncertainty discussed in Sect 3.3. Even the
perfect experiment with infinite statistics and perfect resolution cannot make conclusions about
the size of the corrections. 
This zone is particularly large for the SM-like $h^o$ covering a large fraction of the
LHC ``hole'' (see Sect. 2). On the other hand, to have a decoupling channel can be useful
as it can provide the ``calibration'' of what should be expected for the other channels
without corrections.

In all plots in Figs.~\ref{fig.concorrneg} and~\ref{fig.concorr},
the space above the long dashed line is the plane region that could be accessed 
experimentally
with a modest resolution of 50\% in $O$. Except for the  $h^o$, 
nearly all points with $\tan \beta \gtrsim 20 - 25$
are testable at the one sigma level. The zone above the short dashed line requires an 
experimental resolution of 20\%. Again with the exception of $h^o$, 
this zone covers 
approximately the region 
$\tan \beta \gtrsim 10 - 15$. Therefore, if an experimental resolution of $20\%$ can be achieved at the
Tevatron and LHC, the analysis of the observables proposed in Sect.~3 could be used to search for
indirect signals of SUSY in the main part of the relevant regions 1 and 2 discussed in Sect.~2.

Finally, we estimate the effect of the extra non-decoupling corrections from the SUSY EW sector.
 As discussed in Sect.~3.2, for very large $A_t$ values, the 
SUSY EW corrections do not decouple and this could increase or decrease the 
nondecoupling effect of the SUSY QCD corrections. For the present assumption 
of equally large SUSY parameters and by choosing a particular combination of 
signs that maximizes the size of $ \Delta m_b^{SEW}$, we find that this 
contribution always remains below $ 50 \%$ of the $ \Delta m_b^{SQCD}$ one. 
In order to be conservative, we have performed the exercise of considering this
pessimistic case where the SUSY EW corrections reduce the SUSY QCD signal by $50 \%$ (see Figs.~\ref{fig.concorrnegew} and~\ref{fig.concorrew}).
As in Figs.~\ref{fig.concorrneg} and~\ref{fig.concorr}, the shaded zone in each plot represents the region where the corrections are completely hidden inside the theoretical uncertainty. The long (short) dashed lines again limit the zones where an experimental resolution of $50 \%$ ($20 \%$) is required to achieve a meaningful measurement.




Figures~\ref{fig.concorrnegew} and~\ref{fig.concorrew} show that, as expected, the reachable values of $\tan \beta$
would approximately double the ones in Figs.~\ref{fig.concorrneg} and~\ref{fig.concorr}. Also, the
shaded area has doubled. The points with $\tan \beta \gtrsim 45-55$ are testable
at the one sigma level with a resolution of $ 50 \%$;  $\tan \beta \gtrsim 20-25$ with
a resolution of $20 \%$. In summary, even in this pessimistic case providing 
SUSY EW corrections of considerable size, a large region in the $(m_A, \tan \beta)$ plane remains testable.


\section{Conclusions}

In this paper we have looked for indirect signals of a heavy supersymmetric spectrum 
via its contributions to the radiative corrections to Higgs boson decays. For that purpose we 
have analyzed the dominant SUSY QCD corrections, coming from heavy squarks and gluinos, 
to the Higgs boson decays into quarks. We have studied in detail the nondecoupling contributions, 
especially in the large $\tan\beta$ region where their size is large. In order to search for 
these SUSY signals, we have proposed a set of observables consisting of ratios of Higgs branching 
ratios into third-generation quarks ($H^o,h^o,A^o \to b\bar b$ and $H^+ \to t\bar b$) divided by 
the corresponding ones into third-generation leptons ($H^o,h^o,A^o \to \tau^+ \tau^-$ and 
$H^+ \to \nu \tau^+$). In addition, the observable for top quark decays given by the
ratio of  $B(t \to H^+ b)$ divided by $B(t \to W^+ b)$, complementary to the previous charged Higgs boson
observable in the low $m_{A^o}$ region, has been analyzed. These observables are 
optimal for this purpose 
since the SUSY QCD corrections appear just in the decays to quarks and, therefore, the decays into leptons
can be used  as control channels. 

We have carefully studied any sources of uncertainty that would modify the prediction of the 
observables, previous to the SUSY QCD corrections. In particular, the theoretical uncertainties coming from the 
QCD corrections, and from the errors in the values of the SM parameters involved in the determination of these
observables, 
have been evaluated. In addition, it has been found that the SUSY QCD corrections will allow one to discriminate between 
the MSSM and a nonsupersymmetric 2HDMII, even in the case of a similar Higgs sector mass pattern.

Our detailed study of this set of observables has revealed their high sensitivity to the SUSY QCD
contributions and shown that they are sizable in most of the ($m_{A^o},\tan\beta$) plane. The
corrections to the different observables are strongly correlated. A global analysis of all of them,
together with the experimental determination of $m_{A^o}$ and $\tan\beta$, will be part of the
search for supersymmetric signals at future colliders, especially if the SUSY spectrum turns out to be
very heavy. The measurement of these observables should be the next step after a Higgs boson discovery at 
the LHC or the Tevatron. We have seen that, with a modest experimental resolution of 50\%, the observables 
would show evidence of the SUSY QCD corrections to the $A^o$, $H^o$, and $H^+$ decay widths,
if $\tan \beta \gtrsim 20 - 25$. A better resolution of $20\%$ would show the corrections down to 
$\tan\beta \gtrsim 10 - 15$.
Our study of the SUSY EW effects shows that, even in the pessimistic case where the SUSY QCD corrections get reduced by $50 \%$, an experimental resolution of $20 \%$ would expose the SUSY corrections down to $\tan \beta \gtrsim 20 - 25$.
 Observation of the corrections in the $h^o$ case would require, in general, 
quite large values of $\tan\beta$. We hope our study will encourage our colleagues of CDF, D0, ATLAS, and 
CMS to investigate the actual performance of their detectors.


\section*{Acknowledgments}
This work has been supported in part by
the Spanish Ministerio de Ciencia y Tecnolog{\'\i}a under projects CICYT
FPA 2000-0980 and FPA2000-3172-E. 

  
\begingroup\raggedright\endgroup

\end{document}